\documentclass[12pt,preprint]
{emulateapj}

\usepackage{color,natbib}

\newcommand{\nraoack}{The National Radio Astronomy Observatory is a
facility of the National Science Foundation operated under cooperative
agreement by Associated Universities, Inc.}

\newcommand{\hi}{\ion{H}{1}}

\newcommand{\mc}{\ensuremath{M_{\chi}}}
\newcommand{\sigv}{\ensuremath{\langle\sigma v\rangle_{\chi}}}
\newcommand{\sigvth}{\ensuremath{\langle\sigma v\rangle_{\chi,th}}}
\slugcomment{ApJ, in press}

\shorttitle{Deep Radio Observations of Nearby dSphs}
\shortauthors{Spekkens et al.}

\begin{document}

%\title{A Deep Search for Extended Radio Continuum Emission from the
%Draco Dwarf Spheroidal Galaxy: Implications for Particle Dark Matter}

\title{A Deep Search for Extended Radio Continuum Emission from Dwarf Spheroidal Galaxies: Implications for Particle Dark Matter}

\author{Kristine Spekkens}

\affil{Royal Military College of Canada, Department of Physics, PO Box
17000, Station Forces, Kingston, Ontario, Canada K7K 7B4}

\email{kristine.spekkens@rmc.ca}

\author{Brian S. Mason} 

\affil{National Radio Astronomy Observatory, 520 Edgemont Road,
Charlottesville, VA 22903-2475}

\author{James E. Aguirre}

\affil{University of Pennsylvania, Department of Physics and
Astronomy, 209 South 33rd Street, Philadelphia, PA 19104}

\author{Bang Nhan}

\affil{Department of Astrophysical and Planetary Sciences, University of Colorado,
391 UCB,
Boulder, CO, 80309}

\begin{abstract}
We present deep radio observations of four nearby dwarf spheroidal
(dSph) galaxies, designed to detect extended synchrotron emission resulting from
weakly interacting massive particle (WIMP) dark matter annihilations in
their halos.  Models by Colafrancesco et al.\ (CPU07) predict the
existence of angularly large, smoothly
distributed radio halos in such systems, that stem from electron and
positron annihilation products spiraling in a turbulent magnetic field. We map a total of $40.5 \, {\rm deg^2}$
around the Draco, Ursa Major II, Coma Berenices, and Willman 1 dSphs with the GBT at $1.4 \, {\rm GHz}$ to detect this annihilation
signature, greatly reducing discrete-source confusion using
the NVSS catalog. 
We achieve a sensitivity of $\sigma_{sub} \lesssim 7\,$mJy/beam in our discrete source-subtracted maps, implying that the NVSS is highly effective at removing background sources from GBT maps.  
For Draco we obtained approximately
concurrent VLA observations to quantify the
variability of the discrete source background, and find it to have a
negligible effect on our results. We construct radial surface brightness profiles from each of the subtracted maps, and jackknife the data to quantify the significance of the features therein. At the $\sim10\arcmin$ resolution of our observations, foregrounds contribute a standard deviation of $1.8\,\mathrm{mJy/beam} \leq \sigma_{ast} \leq 5.7\,\mathrm{mJy/beam}$ to our high-latitude maps, with the emission in the Draco and Coma dominated by foregrounds.  On the other hand, we find no significant emission in the Ursa Major II and Willman 1 fields, and explore the
implications of non-detections in these fields for particle
dark matter using the fiducial models of CPU07. For a WIMP mass
$M_\chi = 100\,$GeV annihilating into $b\bar{b}$ final states and $B = 1\,\mu$G, upper limits on the annihilation cross-section for Ursa Major II and Willman I are $\log(\sigv, \mathrm{\,cm^3\,s^{-1}}) \lesssim -25$ for the prefereed set of charged particle propagation parameters adopted by CPU07; this is comparable to that inferred at $\gamma$-ray energies from the two-year Fermi-LAT data.  We discuss three avenues for improving the constraints on \sigv\ presented here, and conclude that deep radio observations of dSphs are highly complementary to indirect WIMP searches at
higher energies.

\end{abstract}


\section{Introduction}
\label{intro}
The standard cosmology derived from astronomical observations such as the cosmic microwave background, the large-scale galaxy distribution, and the kinematics of individual galaxies and clusters predicts that the universal matter density is dominated by dynamically cold, collisionless dark matter \citep[e.g.][]{komatsu10}. Although there are few clues to its nature, its observed abundance requires that a dark matter particle have an annihilation cross section $\sigvth \sim 3 \times 10^{-26}\,\mathrm{cm^3\,s^{-1}}$ if it was once in thermal equilibrium \citep[e.g.][]{porter11}. The correspondence of this cross-section to particles near the weak scale makes weakly interacting massive particles (WIMPs) very attractive dark matter candidates (see \citealt{jungman96}, \citealt{bergstrom00} and \citealt{feng10} for reviews). Whether WIMPs constitute the dark matter, and whether clues to the WIMP identity can be obtained are therefore pressing questions. 

Indirect WIMP searches focus on detecting standard model particles that result from WIMP annihilations or decays. Given the relative paucity of astrophysical sources and the relative robustness of annihilation signature predictions at high energies, $\gamma$-ray searches have yielded some of the tightest constraints on \sigv\ (see \citealt{strigari12} for a recent review). A wide variety of objects have been targeted, including the Galactic center \citep[e.g.][]{abdo10a,dobler10,abazajian12a,abazajian12b,weniger12}, the diffuse Galactic and extragalactic backgrounds \citep[][]{abdo10b,cirelli10,papucci10,baxter11,ackermann12e} and galaxy clusters \citep[][]{ackermann10,abramowski12,ando12,han12}. However, uncertainties in astrophysical backgrounds, the dark matter distributions of the targeted systems and boost factors due to dark matter substructure weaken constraints derived from these studies \citep[e.g.][]{su10,dobler11,inoue11,ackermann12e}.

%. A variety of direct recoil detection experiments are underway, such as DAMA \citep{bernabei08}, CoGeNT \citep{barbeau07}, CDMS+EDELWEISS \citep{ahmed11} and XENON100 \citep{aprile10}. Indirect searches for cosmic rays in the solar neighbourhood (e.g.\ ATIC, \citealt{chang08}; PAMELA,  \citealt{picozza07};  VERITAS, \citealt{weeks02}) and for $\gamma$-rays from more distant astrophysical targets (FERMI, \citealt{atwood09}; MAGIC, \citealt{ferenc05}; WHIPPLE, \citealt{cawley90}) that are produced by WIMP self-annihilations have also been undertaken. Tantalizing signal modulations from direct experiments \citep{aalseth10,bernabei08} and and cosmic ray excesses from  indirect experiments \citep[e.g.][]{chang08,abdo09b,adriani10b}  have been reported. However, conflicts with reported non-detections, potential systematic effects, and difficulties in discriminating particle and astrophysical backgrounds have muddled the interpretation of these signals as the dark matter (see discussions in \citealt{hooper10}, \citealt{porter11}, and references therein).

By contrast, nearby dwarf spheroidal galaxies (dSphs) are attractive targets for indirect dark matter searches: they are strongly dark matter dominated \citep[e.g.][]{mateo98}, and stringent upper limits on their star formation rates imply negligible astrophysical $\gamma$-ray emission \citep[e.g.][]{ackermann12b}. In particular, the line-of-sight integrals of the squared dark matter distributions that are consistent with their stellar kinematics (the ``J-values") predict that the annihilation signal from Draco should be the strongest among the classical dSphs \citep{strigari07,charbonnier11,walker11}. J-values for the ultra-faint dSphs  Ursa Major II (hereafter UMaII), Coma Berenices (hereafter Coma) and Willman 1 (hereafter Will1), while more uncertain, may be a factor of a few larger \citep{strigari08,strigari12}. The basic optical properties of these four dSphs are given in Table~\ref{tab:basic}. 

Accordingly, many $\gamma$-ray experiments have targeted dSphs  to search for WIMP annihilations \citep[e.g.][]{abdo10,ackermann11,aleksic11,aliu12}.  No detections have been reported. The strongest upper limits on \sigv\ for WIMP masses $\mc \lesssim 500\,$GeV stem from observations with the Large Area Telescope (LAT) on Fermi \citep{atwood09}:  analysis of the two-year Fermi-LAT data assuming $\mc = 100\,$GeV and annihilation into $b \bar{b}$ produces upper limits $\sigv \lesssim 10^{-25} \,\mathrm{cm^3\,s^{-1}}$ for individual dSphs, and $\sigv \lesssim 7 \times 10^{-26} \,\mathrm{cm^3\,s^{-1}}$ when observations of 10 dSphs are combined (\citealt{ackermann11}; see also \citealt{geringer-sameth11}). $\gamma$-ray searches are therefore closing in on \sigvth\ expected for a thermal relic at these masses, though sensitivity gains of factors of a few are still required.
%, although an order of magnitude in sensitivity must still be gained before the most interesting thermal relic WIMP models can be tested.

% Models by \citet{strigari08} find that the predicted annihilation signal from Draco among the strongest of the classical dSphs, while \citet{strigari08} show that the fluxes from the ultra-faint dSphs Ursa Major II (hereafter UMaII), Coma Berenices (hereafter Coma) and Willman 1 (hereafter Wil 1)  are among the highest for any Milky Way satellite. The basic properties of these dSphs are given in Table~\ref{tab:basic}. \blue{(elaborate on these dSphs a bit more here.)}
 %Draco \citep[][see Table~\ref{tab:basic}]{baade61a,baade61b} is a particularly good target among dSphs: observations place tight upper limits on its neutral \citep{young99,young00} and ionised \citep{gallagher03} gas content, and deep photometric \citep[e.g.][]{odenkirchen01,cioni05,segall07} and spectroscopic \citep[e.g.][{\it more here}]{walker07} studies place strong constraints on the structure of its dark matter halo relative to other dSphs. Draco is among the most promising candidates for indirect dark matter searches because of its proximity to the Milky Way, and its high, well-modeled dark matter content \citep[][]{strigari07,bringmann09,martinez09,pieri09,kuhlen10,charbonnier11,walker11}. 

A variety of WIMP annihilation channels produce non-thermal electrons and positrons that could be detected when they lose energy through synchrotron, inverse Compton scattering or bremmstrahlung processes \citep[see][for a review]{profumo10}. Compared to $\gamma$-ray searches, this approach is complicated by uncertainties in charged particle propagation and energy losses. Nonetheless, because the expected signals span the electromagnetic spectrum and can be relatively sensitive to \sigv, a variety of multi-wavelength WIMP searches have been carried in the Galactic center \citep[e.g.][]{bergstrom06,hooper07,crocker10,linden10,laha12}, the diffuse background \citep[e.g.][]{hooper08,fornengo12,hooper12} and galaxy clusters \citep[e.g.][]{perez09,colafrancesco06,storm12}. In addition, radio \citep{tasitsiomi04,borriello10,siffert11} and X-ray \citep{jeltema08} observations of Local Group dwarf galaxies have been used to constrain WIMP properties in this context.

%A variety of WIMP annihilation channels produce electrons and positrons \citep[e.g.][]{porter11}, which raises the possibility of detecting synchrotron radiation from their interactions with ambient magnetic fields.  Compared to $\gamma$-ray searches, this approach is complicated by uncertainties in the field strengths and diffusion properties in many sources. Nonetheless, for reasonable assumptions about these properties the predicted signals can be comparatively easier to detect. As such,  cosmic microwave background features \citep[e.g.][]{hooper07,hooper08} and cluster radio halos \citep{colafrancesco06} have both been interpreted in terms of WIMP annihilation signatures. In addition, detected radio emission from the LMC \citep{tasitsiomi04,siffert11} and from M33 \citep{borriello10} have been used to constrain WIMP properties. Searches for dark matter annihilation signatures at radio wavelengths are thus very complementary to those at higher energies from both astrophysical and particle physics viewpoints: the observational systematics and backgrounds differ between the measurements, as do the annihilation products. \blue{Double-check refs in this par.}
%\blue{update this, ref. Profumo+Ullio2010, Jeltema+Profumo2008}. 

\begin{figure}
\epsscale{1.2}
\plotone{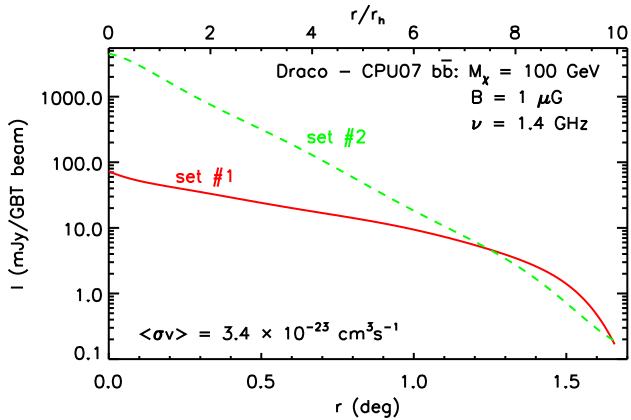}
\caption{Predicted radial surface brightness profiles for Draco by CPU07, scaled to $\nu=1.4\,$GHz and units of mJy per 9.12\arcmin\ beam, appropriate for observations with the GBT at this frequency. The models assume $\mc = 100\,$GeV annihilating into $b\bar{b}$ with $\sigv=3.4 \times 10^{-23}\,\mathrm{cm^3\,s^{-1}}$ (now strongly ruled out by $\gamma$-ray searches), and Milky Way-like diffusion models describing charged particle propagation of the form $D(E) \propto D_0E^\gamma$, with either $D_0 = 3 \times 10^{28}\,\mathrm{cm^2\,s^{-1}}$ and $\gamma = 1/3$ (``set \#1") or  $D_0 = 3 \times 10^{26}\,\mathrm{cm^2\,s^{-1}}$ and $\gamma = -0.6$ (``set \#2"). The upper radial coordinate axis is expressed in units of the Draco half-light radius $r_h$ (Table~\ref{tab:basic}); we apply this model to other dSphs in our sample by scaling it radially to their $r_h$.\\
(A color version of this figure is available in the online journal.)}
\label{fig:pred}
\end{figure}

A series of models by \citet[][hereafter CPU07]{colafrancesco07} suggest that WIMP annihilations in Draco will produce a smoothly distributed, degree-scale radio synchrotron halo. Fig.~\ref{fig:pred} shows the predicted annihilation signal at $\nu=1.4\,$GHz for a representative set of CPU07 models with $\mc = 100\,$GeV annihilating into $b \bar{b}$ and a turbulent magnetic field strength in Draco of $B = 1\,\mu$G. CPU07 adopt Milky Way-like diffusion models describing charged particle propagation of the form $D(E) \propto D_0E^\gamma$, with either $D_0 = 3 \times 10^{28}\,\mathrm{cm^2\,s^{-1}}$ and $\gamma = 1/3$ (``set \#1") or  $D_0 = 3 \times 10^{26}\,\mathrm{cm^2\,s^{-1}}$ and $\gamma = -0.6$ (``set \#2"). Because the size of the diffusion zone corresponds to twice that of Draco's stellar distribution, a crude estimate of the predicted annihilation halos for other dSphs is obtained by scaling the profiles radially according to their half-light radii $r_h$ (Table~\ref{tab:basic}). This approach assumes a fixed annihilation flux, 
%dSphs in Fig.~\ref{fig:pred} is valid for all dSphs and the WIMP properties adopted by CPU07: 
 which is reasonable given the similarity between the J-values of the systems considered here \citep{strigari08,strigari12}. As discussed in \S\ref{discuss:cpu07:limits}, ``set \#2" is likely to be most appropriate for dSphs, with the model assumptions producing order-of-magnitude estimates of the predicted annihilation flux for a given \sigv.
   
 Fig.~\ref{fig:pred} plots the predicted halo intensities for CPU07's optimistic choice of $\sigv \sim 3.4 \times 10^{-23}\,\mathrm{cm^{3}s^{-1}}$, a value that is now strongly ruled out by $\gamma$-ray searches. With this \sigv, the predicted signals would be easily detectable with existing single-dish radio telescopes.  
While CPU07's \sigv\ is no longer relevant, realistic values of \sigv\ can be probed with the models of Fig.~\ref{fig:pred} by scaling them linearly in intensity when comparing to observations. 

%In Fig.~\ref{fig:pred}, we show the predicted surface brightness distribution from the fiducial CPU07 models at an observing frequency $\nu = 1.4\,$GHz (by requiring that the profile integrate to the total flux at that frequency in their fig.~12 left), and we plot the radial extent of the halo in units of the Draco half-light radius of $10\,$\arcmin\ determined by \citet{martin08}. While the CPU07 models are specific to Draco, we will adapt them to other dSphs by adopting the CPU07 dark matter halo model (and hence the same surface brightness scaling as in CPU07) and by scaling the predicted halo radially according to $r_h$ in Table~\ref{tab:basic}.  

 However, few strong constraints on the radio flux densities of dSphs exist \citep[e.g.][]{fomalont79}, and no dedicated searches for extended radio halos in dSphs have been performed\footnote{Note that the limit for Draco from \citet{fomalont79} shown in fig.~12 of CPU07 has little relevance in the context of dark matter searches: these interferometric observations resolve out structures larger than few arcminutes and are therefore insensitive to degree-scale annihilation halos.}.  The dearth of extended emission searches in dSphs stems in part from the observational challenges associated with reaching sensitivities below the confusion limit of single-dish telescopes, imposed by discrete background sources \citep[e.g.][]{condon74}.  Radio searches for WIMPs in dSphs therefore require a combination of single-dish observations to detect the predicted extended emission, and higher resolution interferometric observations to subtract discrete sources from the single-dish maps and beat their nominal confusion limit. 
%Relative to gas-rich dIrrs such as M33 and the LMC, the unconstrained magnetic field strengths in dSphs complicate the interpretation of the resulting radio data (see \S\ref{discuss}). Nonetheless, the higher dark matter content of many dSphs compared to the dIrrs \citep[e.g.][]{mateo98} and the ``cleaner" putative signal makes them attractive targets.

Accordingly, we have obtained deep radio continuum maps of dSphs using the Robert C. Byrd Green Bank Telescope (GBT) in order to search for extended radio halos resulting from WIMP annihilations. In this paper, we present $\nu=1.4\,$GHz observations of degree-scale regions centred on Draco, UMaII, Coma and Will1.  We subtract discrete sources from the Stokes I maps using the Karl G. Jansky Very Large Array (VLA) Sky Survey (NVSS; \citealt{condon98}) to reduce discrete-source confusion (\S\ref{data:gbt}~and~\S\ref{halo:prof}), and confirm that discrete source variability is unimportant to our analysis with near-simultaneous VLA observations of the Draco field (\S\ref{data:vla}~and~\S\ref{halo:var}). We find that our observations are foreground-limited (\S\ref{data:gbt}~and~\S\ref{discuss:foregrounds}), and that foreground contamination precludes a detailed analysis of the Draco and Coma fields (\S\ref{discuss:cpu07:draco}). However, after accounting for the filtering of large-scale flux due to our baselining procedure (\S\ref{halo:bias}), we use non-detections in the UMaII and Will1 fields to place upper limits on \sigv\ in the context of the models shown in Fig.~\ref{fig:pred} (\S\ref{discuss:cpu07:limits}). We demonstrate that for a class of dark matter models like those of CPU07, deep radio observations are highly complementary to $\gamma$-ray searches for constraining the properties of particle dark matter (\S\ref{discuss:cpu07:limits}). We discuss the limitations of our analysis due to foregrounds (\S\ref{discuss:foregrounds}) and plausible magnetic field strengths in dSphs (\S\ref{discuss:bfields}), and describe an observational program to improve the limits on \sigv\ found here (\S\ref{discuss:future}).

\begin{deluxetable*}{lcrcc}
\tablecaption{Basic Properties of the dSph Sample \label{tab:basic}}
\tablehead{ \colhead{Target} & \colhead{$(\alpha_0,\delta_0)$}  & \colhead{$(l_0,b_0)$} & \colhead{$r_h$}        & \colhead{Reference} \\
                                               &  \colhead{(J2000)}             &                         & \colhead{(arcmin)}   &   \\
                    \colhead{(1)}       & \colhead{(2)}                      & \colhead{(3)}  &  \colhead{(4)}            & \colhead{(5)}              }

\startdata
Draco  & $17^\mathrm{h}20^\mathrm{m}14.4^\mathrm{s} \pm 0.6^\mathrm{s}$, 57\arcdeg54\arcmin54\arcsec $\pm$ 08\arcsec & 86.4\arcdeg, 34.7\arcdeg &     $10.0^{+0.3}_{-0.2}$      &  1 \\
UMaII  & $08^\mathrm{h}51^\mathrm{m}29.9^\mathrm{s} \pm 4.0^\mathrm{s}$, 63\arcdeg07\arcmin59\arcsec $\pm$ 07\arcsec &  152.5\arcdeg, 37.4\arcdeg  &     $14.1 \pm 0.3$            &  2 \\
Coma  & $12^\mathrm{h}26^\mathrm{m}59.0^\mathrm{s} \pm 0.9^\mathrm{s}$, 23\arcdeg54\arcmin27\arcsec $\pm$ 08\arcsec &  241.9\arcdeg, 83.6\arcdeg  &     $5.8 \pm 0.3$ &  2 \\
Will1     & $10^\mathrm{h}49^\mathrm{m}21.9^\mathrm{s} \pm 0.8^\mathrm{s}$, 51\arcdeg03\arcmin10\arcsec $\pm$ 11\arcsec &  158.6\arcdeg, 56.8\arcdeg   &     $2.3^{+0.2}_{-0.4}$    & 1 
\enddata
\tablecomments{ Col.\ 2: RA and dec $(\alpha_0, \delta_0)$ of the stellar dSph centroid;  Col.\ 3: Galactic coordinates $(l_0,b_0)$ computed from $(\alpha_0, \delta_0)$; Col.\ 4: half-light radius of an exponential model of the stellar distribution; Col.\ 5: Reference: (1) \citet{martin08}; (2) \citet{munoz10}. }
\end{deluxetable*}

% converted coords using: http://www.robertmartinayers.org/tools/coordinates.html
%Draco, deg: 260.06, 57.915
%UMaII, deg: 132.8746, 63.1331
%Com: 186.7458, 23.9076
%Wil1: 162.3412, 51.0528

\section{Observations and Data Processing}
\label{data}

In order to search for extended radio halos in the dSphs in Table~\ref{tab:basic}, we mapped $1.5^\circ$- to $4^\circ$-square regions centered on each dSph with the GBT at a frequency of $\nu=1.4\,$GHz. 
%This mapping frequency is a compromise between
%synchrotron signal strength and mapping speed (which
%decrease/increase with increasing frequency, respectively). 
This frequency affords the use of publicly available NVSS
survey data to subtract discrete background sources. Below, we discuss the
details of our GBT observations (\S\ref{data:gbt}) as well as our near-simultaneous VLA observations of the Draco field
to assess the impact of discrete source variability in our final GBT maps (\S\ref{data:vla}).

\subsection{GBT Observations and Data Processing}
\label{data:gbt}

Observations were made with the GBT in numerous observing sessions
under the auspices of programs AGBT07C085 (for Draco) and AGBT09A085
(for UMaII, Coma, and Will1).  Maps were made in the on-the-fly mode, scanning at 104\arcmin/minute, with 2 seconds of
integration time per pixel, giving a total time for one $4\arcdeg
\times 4\arcdeg$ map of 2.8 hours.  The scan rate was chosen so that
with the spectrometer backend integration time of 1 second, the
telescope motion is 20\% of a beam full width at half maximum (FWHM) per integration, resulting in
negligible beam smearing.  RA and dec scans were interleaved to reduce
the effect of $1/f$ noise, either atmospheric or instrumental, on the
final map.  The GBT backends were configured to simultaneously record the receiver flux for all polarization products using both the Digital Continuum Receiver (DCR) and the GBT Spectrometer.  The spectrometer produces both continuum and \hi\ information.  We defer a discussion of polarization and spectral data products to a future publication (J. Aguirre et al. 2013, in preparation).  In this paper, we restrict our analysis to the two DCR XX and YY polarization channels, which, when averaged over parallactic angle (PA), approximate the continuum Stokes I intensity. The basic characteristics of the observations are presented in Table~\ref{tab:obs}.
 
 In order to search for extended halos for realistic values of \sigv, we need to reach sensitivities below the confusion limit of the GBT at $\nu = 1.4\,$GHz by subtracting discrete background sources (see \S\ref{intro}). The $45\arcsec$-resolution NVSS provides a strong
handle on this background. Note that although FIRST \citep[Faint Images of the Radio Sky at Twenty Centimeters;][]{becker95} boasts a higher
angular resolution and sensitivity per beam than the NVSS, it has the
significant disadvantage of resolving out smoothly distributed
emission on angular scales larger than $\sim1\arcmin$. 
%We also simulated the impact of  subtracting sources fainter than the detection threshold of the NVSS
%  catalog using the number densities reported in \citet{schinnerer04},
%  and find that this population contributes negligibly to the total
%  map flux. 
  We therefore subtract only sources
  detected by the NVSS. The GBT is
sufficiently sensitive that NVSS sources are detected at high
signal-to-noise in each individual $\sim 10$-second raster scan
across a given map.  In order to reduce cross-calibration
uncertainties and to obtain a high-quality subtraction of the discrete
sources using the NVSS, we directly calibrate the GBT data off of the
latter.

\begin{figure}
\epsscale{1.2}
\plotone{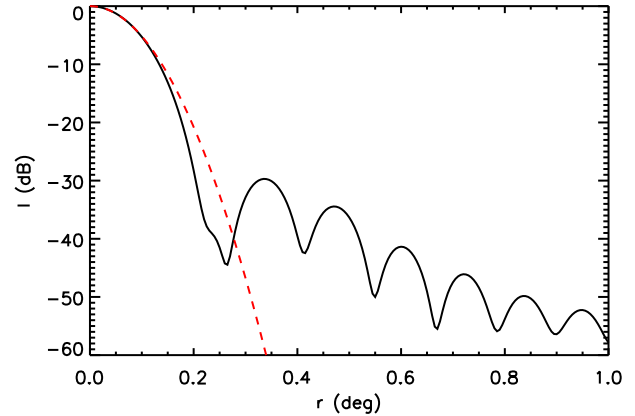}
\caption{GBT beam model (solid line) at $\nu = 1.4\,$GHz used to convolve NVSS maps of each field to the GBT resolution in order to subtract discrete sources from each dSph field. Above a response of  $I= -7\,$dB, the GBT beam is well-approximated by a Gaussian with a FWHM of 9.12\arcmin\ (dashed line).\\
(A color version of this figure is available in the online journal.)}
\label{fig:beams}
\end{figure}

For each scan, the raw GBT time-ordered data $d(t_i)=d_i$ are modeled
as
\begin{equation}
d_i = p_i + s \, NVSS_i \,\,\,,
\label{eq:red1}
\end{equation} 
where $p_i$ is a first-order polynomial (mean and slope) to remove baseline drifts in the data, $NVSS_i$ is the NVSS template (in units of mJy
per GBT beam) convolved to the GBT resolution, and $s$ is a scale factor determined for each scan. Beam-maps constructed from scans across bright continuum sources show that the $\nu = 1.4\,$GHz GBT beam is well-approximated by a Gaussian with a FWHM of 9.12\arcmin\ above $-7\,$dB, but that it falls off more rapidly than this at fainter levels. Our beam-maps are in good agreement with the GBT ray tracing model of Sri Srikant (priv. comm.); we therefore use an azimuthal average of this model, shown in Fig.~\ref{fig:beams}, to convolve the NVSS data to the GBT resolution. Use of this more accurate model instead of a Gaussian significantly improves point source subtraction.

  The values of $p_i$ and $s$ are determined
by a least squares fit to the given scan's data. We use a first-order polynomial for $p_i$ as a compromise between removing 1/f noise and not removing too much large-scale emission, and place quantitative limits on our sensitivity to extended halos in \S\ref{halo:bias}. The baseline-subtracted, NVSS-calibrated data $D_i$ are then given by
\begin{equation}
D_i = (d_i - p_i)/s \,\,\,.
\label{eq:red2}
\end{equation}
For conciseness, we refer to maps created from the $D_i$ as ``unsubtracted" because they contain discrete sources. By contrast, the baseline-subtracted, NVSS-calibrated and discrete source-subtracted data $D_{sub,i}$ (which we will call ``subtracted") that we require for our analysis are given by
\begin{equation}
D_{sub,i} = (d_i - p_i - s \, NVSS_i)/s \,\,\,.
\label{eq:red3}
\end{equation}

\begin{figure*}
\epsscale{1.2}
\plotone{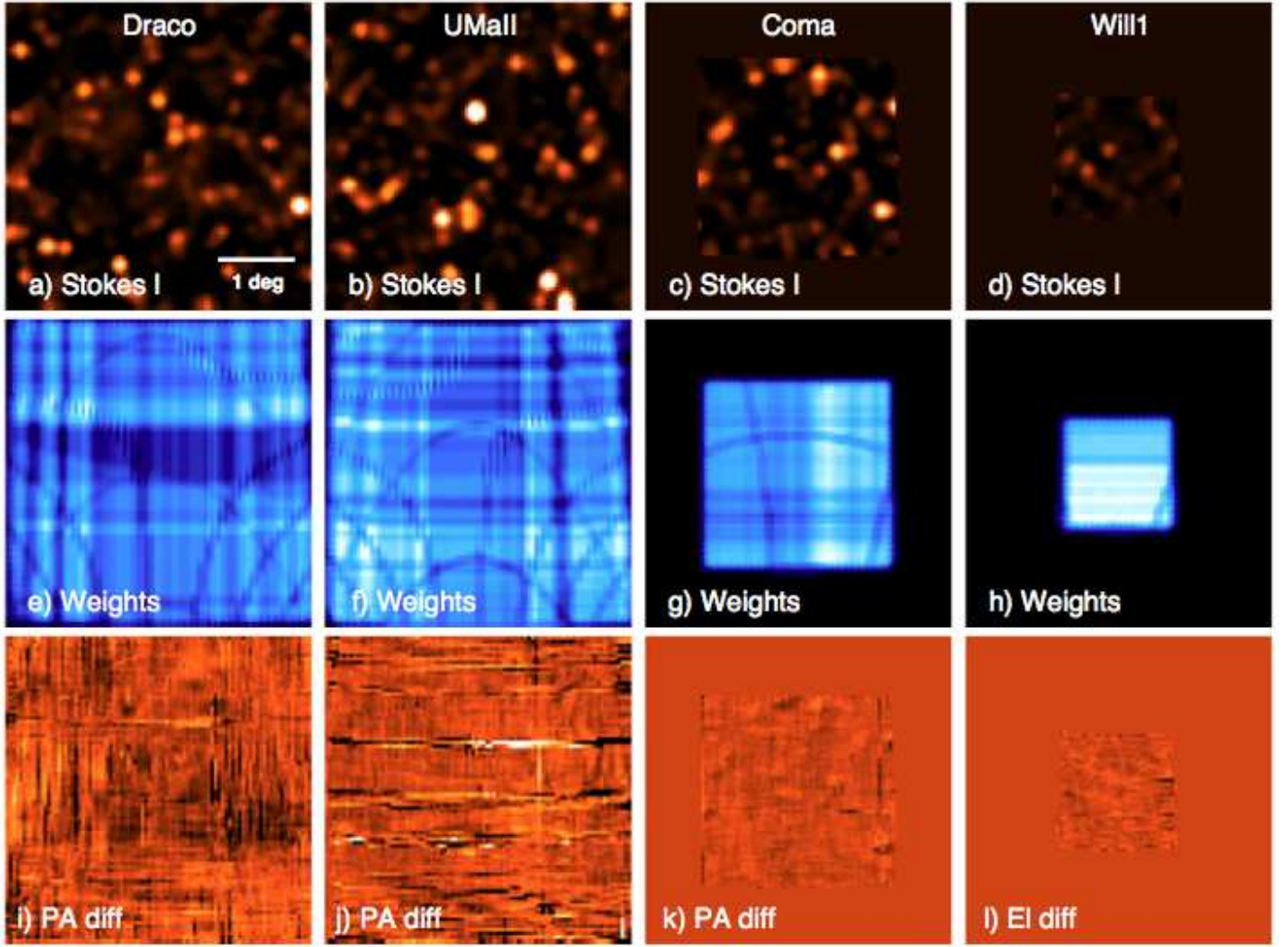}
\caption{GBT data products. Each column corresponds to a distinct field that is labeled in the top row. All panels span $4\arcdeg \times 4\arcdeg$: the horizontal line in the bottom-right corner of a) shows an angular scale of $1\arcdeg$, and applies to all panels. The borders around the Coma and Will1 fields in the third and fourth columns are blanks around these smaller maps (see Table~\ref{tab:obs}) . {\it Top row:} Baseline-subtracted, NVSS-calibrated ``unsubtracted" Stokes I maps  of a) Draco, b) UMaII, c) Coma, d) Will1.  The linear intensity scale ranges from -10 to 250 mJy/beam. {\it Middle row:} Weight maps for the Stokes I maps in the row above. The linear colorscale ranges from 0 (black) to 1 (white).  {\it Bottom row:} Difference map, created by jacknifing the data, with the largest standard deviation for each field: i), j), k) PA, and l) elevation. The linear intensity scale ranges from -10 to 25 mJy/beam. \\
(A color version of this figure is available in the online journal.)}
\label{fig:data}
\end{figure*}

\begin{figure}
\epsscale{0.75}
\plotone{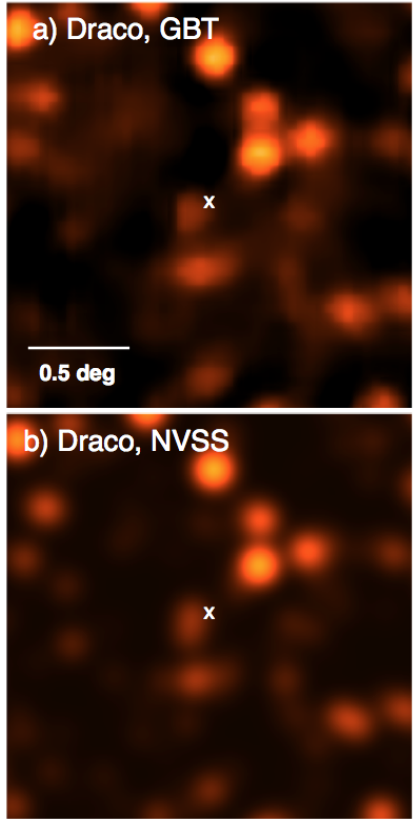}
\caption{Comparison of the inner $2\arcdeg \times 2\arcdeg$ of the Draco field from a) our GBT observations and b) the NVSS, convolved to the GBT resolution. The horizontal line in the bottom-left corner of a) shows an angular scale of $0.5\arcdeg$.  In both panels, the linear intensity scale ranges from -10 to 250 mJy/beam and the cross denotes the stellar centroid of Draco (Table~\ref{tab:basic}).\\
(A color version of this figure is available in the online journal.)}
\label{fig:nosub}
\end{figure}

\begin{figure}
\epsscale{0.8}
\plotone{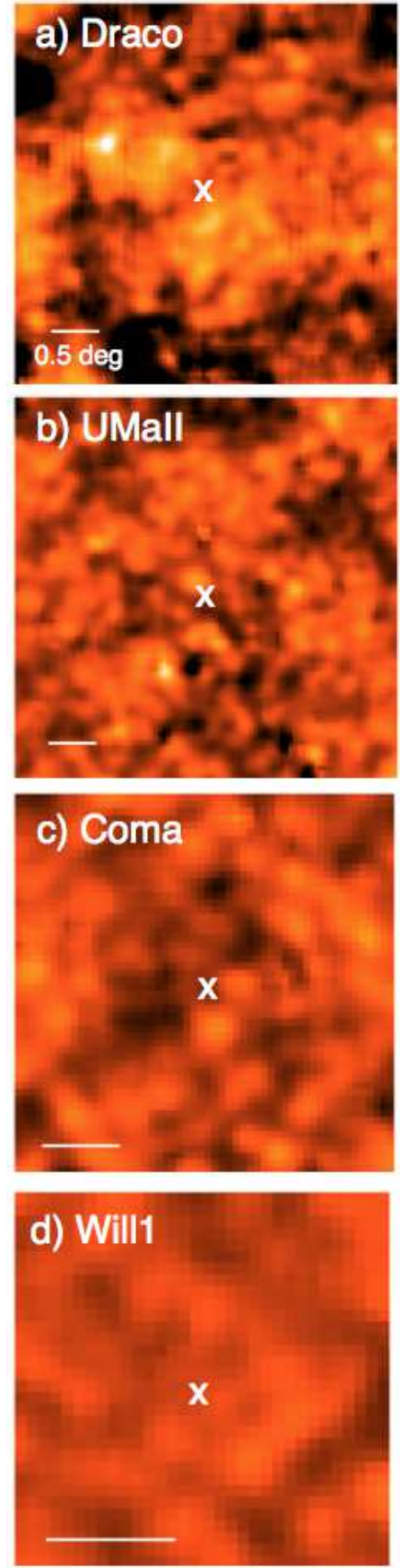}
\caption{Discrete-source subtracted Stokes I maps of a) Draco, b) UMaII, c) Coma, d) Will1.  The linear intensity scale ranges from -10 to 25
  mJy/beam: note that this upper limit is a factor of 10 smaller than
  that in Fig.~\ref{fig:data}a-\ref{fig:data}d. The horizontal line in the lower left corner of each panel is 0.5\arcdeg\ in length, and the cross denotes the optical centroid of each dSph (Table~\ref{tab:basic}).\\
  (A color version of this figure is available in the online journal.)}
\label{fig:sub}
\end{figure}

The subtracted scans were visually
inspected for dramatic deviations from zero which could result from
interference, atmospheric fluctuations due to poor weather conditions, or other problems; a small fraction
of scans were rejected for these reasons. We also discard scans obtained at elevations below 15\arcdeg, which is the case only in the Draco field. The remaining data had XX and YY polarizations combined to produce Stokes I, and gridded onto the sky
using a standard cloud-in-cell method on a 40\arcsec\ grid. The maps intended for visual inspection, and presented in Figs.~\ref{fig:data}--\ref{fig:sub}, were constructed using a median gridding kernel in which the median of the integrations within a $3\times3$ pixel box centered on each pixel is adopted as the final value. However, all computations were performed on maps constructed without median gridding, to ensure that the map weights were linearly propagated throughout the analysis. 
%The observing and map parameters are given in the second column of Table~\ref{tab:obs}.

The unsubtracted Stokes I GBT maps obtained from $D_i$ for the four mapped fields are shown in the top row of Fig.~\ref{fig:data}. All fields are plotted on the same angular scale, and are dominated by discrete sources. This is illustrated in Fig.~\ref{fig:nosub} for Draco, which shows the inner $2\arcdeg \times 2\arcdeg$ of the unsubtracted Stokes I GBT map and the NVSS map of the same region, convolved to the GBT resolution.  The striking similarity between the discrete source patterns in the two panels suggests that source variability does not strongly affect our subtracted maps; we confirm this using near-simultaneous VLA observations in \S\ref{halo:var}. 

The second row in Fig.~\ref{fig:data} shows the weight maps for the Stokes I data, which quantify the relative data contribution to each pixel on a linear scale from 0 to 1 (black and white in Fig.~\ref{fig:data}e-\ref{fig:data}h, respectively). 
%The arcs in the weight maps reflect the transformation between the azimuthal scanning coordinates and the equatorial sky coordinates. 
All calculations performed include only map pixels whose weights exceed half the median value of the non-zero map weights. We verified that our results do not depend on the precise value of the median weight threshold adopted. 

 The third row in Fig.~\ref{fig:data} shows ``difference maps" for each field. The difference maps are a measure of mapping artifacts, which dominate the uncertainties in our analysis (see below). Specifically, we jacknife the data to produce three pairs of maps for each field created from only half of the collected data: that below (``Split A") and above (``Split B") the median time of observation, the median telescope elevation and the median telescope PA. The morphology of any real sky feature should be independent of these observing properties. Accordingly, for a perfectly mapped region Split A and Split B for each property would be identical, and subtracting Split B from Split A would produce a map containing no sky signal. We define our difference map as (Split B - Split A)/2, so that the variance in these maps has the same scaling as that of Stokes I (because halves of the data are subtracted to produce the difference map). Note that finer data splits (e.g.\ into thirds) were also attempted, but the resulting maps were too noisy to be useful. 
 
 We therefore use the statistics of the features in the observed difference maps as a measure of the mapping errors. We create difference maps in time, elevation and PA for each field. In Fig.~\ref{fig:data}i-\ref{fig:data}l, we show the difference map for each field where the standard deviation of the pixels is the largest, and use this map for our error analysis in \S\ref{halo:prof}. For each field, the telescope property jacknifed to produce the plotted difference map is given in the bottom-left corner of the corresponding panel in Fig.~\ref{fig:data}: PA for Draco, UMaII and Coma, and elevation for Will1. As expected, the difference maps have more gaps than the Stokes I maps of Fig.~\ref{fig:data}a-\ref{fig:data}d,  because a pixel must have data in each jacknifed half to be included in the difference map.
   
 The subtracted Stokes I GBT maps obtained from $D_{sub,i}$ for the four mapped fields are shown in Fig.~\ref{fig:sub}, and form the basis of our search for extended radio emission from the dSphs at each field center.  Note that the upper colorscale limit in this
figure an order of magnitude smaller than in
Fig.~\ref{fig:data}a-\ref{fig:data}d: the vast majority of the detected flux in the unsubtracted maps stems from the discrete source population. 
Positive or negative features in the subtracted maps represent enhancements or depressions relative to the (positive) absolute sky brightness, which was subtracted during processing by virtue of $p_i$ in equation~(\ref{eq:red1}). 
%The peak flux in any of the subtracted maps is $26\,$mJy/GBT
%beam and appears to stem from an unresolved source at
%$(17^\mathrm{h}27^\mathrm{m}40^\mathrm{s},\,58^\circ25'$) in the Draco
%field; the second-highest peak is $21\,$mJy/GBT
%beam at $(8^\mathrm{h}54^\mathrm{m}34^\mathrm{s},\,62^\circ17')$ in the
%UMaII field. It is also unresolved. Neither position coincides with an
%NVSS source, suggesting that the emission stems from the Galactic foreground.  The Draco residual appears to be embedded in a ring-like
%structure centered on $(17^\mathrm{h}25^\mathrm{m},\,58^\circ21"$), which corroborates this hypothesis.

There are no detectable residuals at the locations of the bright NVSS
sources that were subtracted from the maps in Fig.~\ref{fig:sub}, nor is there a correlation
between the residual map emission and NVSS source locations. As a check, we repeat the analysis described in \S\ref{halo} by blanking regions within 1.5 GBT beams of bright NVSS sources: our results remain unchanged. Indeed, there are no discernible discrete source artifacts in our subtracted maps at all. This demonstrates that the NVSS is highly effective at reducing source confusion at the GBT resolution.

We quantify the sensitivity of the unsubtracted and subtracted GBT maps of each field in Table~\ref{tab:noise}. As expected, the standard deviation $\sigma_{usub}$ (col.\ 2) of the pixels in the unsubtracted Stokes I maps well exceeds the standard deviation $\sigma_{sub}\lesssim 7\,$mJy/beam (col.\ 3) in the subtracted maps.  In turn, the values of $\sigma_{sub}$ exceed the thermal noise expected from the radiometer equation by about an order of magnitude, as a result of both mapping errors and real sky variations. To disentangle the two, we use the statistics of the difference maps in Fig.~\ref{fig:data}i-\ref{fig:data}l. The estimated contribution of mapping uncertainties to the subtracted maps is $\sigma_{map}$ (col.\ 4), and corresponds to the standard deviation of the pixels in the difference maps of Fig.~\ref{fig:data}i-\ref{fig:data}l. Assuming that the mapping and sky variances are additive, $\sigma_{ast}$ (col. 5) is then the contribution to $\sigma_{sub}$ from astrophysical sources. We discuss the interpretation of $\sigma_{ast}$ as a measure of Galactic foregrounds in \S\ref{discuss:foregrounds}: in this context, Table~\ref{tab:noise} shows that our sensitivity is foreground-limited.
 
The final column in Table~\ref{tab:noise} is a measure of the dynamic range of the GBT observations, which we define as the ratio $DR$ of the peak flux in the {\it unsubtracted} maps to the standard deviation $\sigma_{sub}$ of the {\it subtracted} maps: $DR$ is thus an indication of our ability to probe faint structures in the field after discrete source subtraction. We find $DR \sim 140$ for the UMaII field, and note that the lower values for the other fields arise simply because the brightest continuum sources therein are fainter. Because we find no discrete source artifacts in any of the fields, we conclude that $DR$ is only a lower limit to the dynamic range achievable for deep continuum observations with the GBT. Our discrete source subtraction is not an important contributor to the noise in the subtracted maps. It is therefore feasible to produce degree-scale, discrete-source subtracted GBT maps with sensitivities $\sigma_{sub}\lesssim 7\,$mJy/beam at $1.4\,$GHz, well below the nominal confusion limit.

% THE OLD TABLE (pre27jun12)
%\begin{deluxetable}{lcc}
%\tablecaption{Observing and Map Properties \label{tab:obs}}
%\tablehead{ \colhead{Property} & \colhead{GBT data} & \colhead{VLA data}}
%\startdata
%Observation date               &     Oct -- Dec 2007                &    4 Nov 2007                  \\
%Time on source                  &       8 hours          &      5.4 hours                \\
%Map dimensions                &      $4^\circ \times 4^\circ$                &     $3^\circ \times 4^\circ$                 \\
%Map centre (J2000)         &    $\mathrm{17^h20^m,\,\,57^\circ54'}$                   &       $\mathrm{17^h18^m,\,\,57^\circ53'}$                \\
%Resolution (FWHM)           &    $9.12\arcmin \times 9.12 \arcmin$                    &      $6.8\arcsec \times 5.3\arcsec\,@83^\circ$                 \\
%RMS sensitivity                  &         FILL IN                &     $0.23\,$mJy/beam                \\
%\enddata
%\end{deluxetable}

% THE NEW TABLE WITH ALL FOUR dSphs
\begin{deluxetable*}{lccccc}
\tablecaption{Observation and Map Properties \label{tab:obs}}
\tablehead{\colhead{Field} & \colhead{Observing Dates} & \colhead{Integ.Time} & \colhead{Map Centre} & \colhead{Dimensions} & \colhead{Resolution}  \\
                     \colhead{(1)}        &   \colhead{(2)}            &  \colhead{(3)}              &  \colhead{(4)}                 &  \colhead{(5)}                &  \colhead{(6)}               }         
\startdata
Draco (GBT) & 2007 October -- December\tablenotemark{a}  & $14.8\, {\rm h}$  &  $\mathrm{17^h20^m,\,\,57^\circ55'}$  &  $4^\circ \times 4^\circ$     &  $9.12\arcmin \times 9.12 \arcmin$    \\
UMaII (GBT) & 2009 February -- March\tablenotemark{b} & $18.8\, {\rm h}$  &  $\mathrm{8^h52^m,\,\,63^\circ08'}$  &  $4^\circ \times 4^\circ$     &  $9.12\arcmin \times 9.12 \arcmin$    \\
Coma (GBT) & 2009 February -- March & $8.6\, {\rm h}$  &  $\mathrm{12^h27^m,\,\,23^\circ54'}$  &  $2^\circ.5 \times 2^\circ.5$     &  $9.12\arcmin \times 9.12 \arcmin$   \\
Will1 (GBT) & 2009 February\tablenotemark{c} & $1.8\, {\rm h}$  &  $\mathrm{10^h49^m,\,\,51^\circ03'}$  &  $1^\circ.5 \times 1^\circ.5$     &  $9.12\arcmin \times 9.12 \arcmin$   \\
Draco (VLA) & 2007 November 4      & $5.4 \, {\rm h}$   &  $\mathrm{17^h18^m,\,\,57^\circ53'}$     &  $3^\circ \times 4^\circ$ &  $6.8\arcsec \times 5.3\arcsec$   
\enddata
\tablecomments{Col.\ 1: Field name. Instrument used to obtain the observations described in remaining columns is given in parentheses. Col.\ 2: Dates when majority of data were acquired. Col.\ 3: Total integration time for field. Col.\ 4: Map centre. Col.\ 5: Final map dimensions. Col.\ 6: Angular resolution of final map. }
\tablenotetext{a}{$81\%$ of data acquired during 2007 October -- December; $19\%$ acquired during make-up sessions in 2009 and 2010.}
\tablenotetext{b}{$91\%$ of data acquired during 2009 February -- March; $9\%$ acquired during make-up sessions in 2010 June.}
\tablenotetext{c}{$74\%$ of data acquired during 2009 February; $26\%$ acquired during make-up sessions in 2010 June.}
\end{deluxetable*}

\begin{deluxetable}{lccccc}
\tablecaption{Noise Properties of the GBT maps \label{tab:noise}}
\tablehead{\colhead{Field}   & \colhead{$\sigma_{usub}$} & \colhead{$\sigma_{sub}$}  & \colhead{$\sigma_{map}$} & \colhead{$\sigma_{ast}$} & \colhead{$DR$}\\
                                               &  \colhead{(mJy/bm)}              & \colhead{(mJy/bm)}                &  \colhead{(mJy/bm)}       & \colhead{(mJy/bm)}     & \\
                     \colhead{(1)}        &   \colhead{(2)}            &  \colhead{(3)}              &  \colhead{(4)}                 &  \colhead{(5)}              & \colhead{(6)}  
}
\startdata
Draco   & 33 & 6.6 & 3.4   & 5.7  &    88 \\
UMaII     & 50 & 6.3 & 5.1   & 3.7  &   142 \\
Coma   & 34 & 3.6 & 1.3   & 3.3   &    87  \\
Will1     & 14 & 2.3 & 1.5   &  1.8  &   37  
\enddata
\tablecomments{Col.\ 1: Field name. Col.\ 2:  standard deviation of pixels in unsubtracted map. Col.\ 3: standard deviation of pixels in subtracted map.  Col.\ 4: estimated contribution to $\sigma_{sub}$ in col.\ 3 from mapping uncertainties, or the standard deviation of the pixels in the difference map in Fig.~\ref{fig:data}i-\ref{fig:data}l. Col.\ 5: estimated contribution to $\sigma_{sub}$ from astrophysical sources: $\sigma_{ast}^2 = \sigma_{sub}^2 - \sigma_{map}^2$. Col.\ 6: Dynamic range: ratio of peak brightness in unsubtracted map and $\sigma_{sub}$. }
\end{deluxetable}

\subsection{VLA Observations and Data Processing}
\label{data:vla}

To constrain the variability of the discrete sources in the Draco field, we obtained VLA observations under program AA315 on 2007 November 4,  near the middle of the 3-month period over which the GBT data for this field were taken.  The VLA was in B-configuration during that observing cycle, and we therefore configured the observing setup to match that of the FIRST\ survey, with one 7-channel,  $3\,$MHz, dual-polarization frequency band centered at each of $1365\,$MHz and $1465\,$MHz.  A comparison of the measured fluxes from our observations to the corresponding FIRST\ catalog entries provides a conservative upper limit on the variability in the NVSS sources in the same field, which we used for discrete source subtraction of the GBT data (see \S\ref{halo:var}). A total of 5.4 hours were spent observing 110 snapshot pointings with the VLA in the Draco field, with flux calibrators observed at the start, middle and end of the observations and phase calibrators observed every 15-30 mins. 

The mapping scheme was designed to reproduce the FIRST\ pointing grid in a $4^\circ \times 4^\circ$ region centered on Draco,  but an error in the interim VLA mapping software produced smaller pointing offsets than requested in RA. The final map therefore spans  $17^\mathrm{h}05^\mathrm{m} 30^\mathrm{s}< \alpha < 17^\mathrm{h}28^\mathrm{m}30^\mathrm{s}$ and $55^\circ 50\arcmin < \delta < 59^\circ \,50\arcmin$: this $3^\circ \times 4^\circ$ region is centered on Draco in dec, but offset from it by $\sim 0.5^\circ$ to the West. The basic characteristics  of the observations are presented in Table~\ref{tab:obs}.

The VLA data were reduced in AIPS \citep{greisen03}. After interactively flagging bad baselines due to interference or poor instrumental performance, the flux calibrator was used to correct for the bandpass response functions of the antennas. The data were then flux and phase calibrated using standard AIPS routines, and the visibilities at $1365\,$MHz and $1465\,$MHz averaged. Each pointing was separately imaged and cleaned down to a residual noise level of $0.6\,$mJy, and then corrected for the geometric distortion produced by the VLA's snapshot mode. The pointings were then mosaicked into a single, primary beam-corrected image. 
%The average RMS noise level of the map is $\sigma=0.23\,$mJy, which rises towards the map edge. The poorer resolution and sensitivity of our map compared to FIRST (FWHM$_{F} = 5.4\arcsec$, $\sigma_{F} = 0.14\,$mJy; \citealt{becker95}) is likely caused by a combination of edge and cleaning effects, our use of the interim VLA, as well as differences between the calibration adopted here and the FIRST pipeline scripts. 

Source fluxes in the VLA map were measured by fitting Gaussian components using the AIPS task JMFIT. To compare these measured fluxes to the corresponding FIRST catalog entries, we follow the procedure of \citet{deVries04}: a) we cross-correlate the positions of our measured sources with those of the 08Jul16  FIRST catalog using a matching radius of $3\arcsec$ (roughly half a synthesized beam), b) we compare the measured peak fluxes of unresolved, matched sources to that of their FIRST counterparts, restricting the comparison to sources brighter than 2\,mJy  (c.f.\ fig.~2 of \citealt{deVries04}). 
%Note that differences in the restoring beam as well as in cleaning and flux measurement algorithms preclude a comparison of integrated fluxes or resolved sources. 
A total of 116 sources meet these criteria; we examined each one individually to ensure that none were unresolved components of a larger ``parent" system.

\begin{figure}
\epsscale{1.1}
\plotone{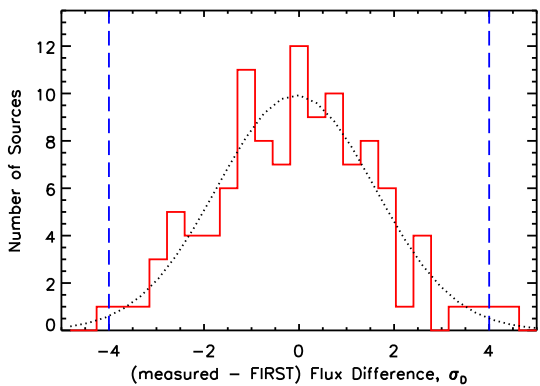}
\caption{Difference between measured peak fluxes of unresolved sources in the Draco VLA field and FIRST catalog peak fluxes, expressed in units of the net measurement uncertainty $\sigma_D$. The dotted line shows the best-fitting Gaussian to the distribution, and the vertical dashed lines denote our variability threshold of $4\sigma_D$. \\ (A color version of this figure is available in the online journal.)}
\label{fig:VLAvar}
\end{figure}

Fig.~\ref{fig:VLAvar} shows the distribution of (measured - FIRST)  peak flux differences for the 116 sources. The flux differences are expressed in units of the net statistical uncertainty:   $\sigma_D = \sqrt{\sigma_{m}^2 + \sigma_{F}^2}$, where $\sigma_{m}$ is the uncertainty returned by JMFIT\  and $\sigma_{F}$ is the RMS map noise at the source location listed in the FIRST\ catalog. Although we applied the FIRST clean bias correction to our measured fluxes \citep{white97}, differences in absolute calibration and cleaning between the catalogs produced a non-zero median flux difference of  $0.56\,$mJy. This has been subtracted from Fig.~\ref{fig:VLAvar}. The dotted line in the figure shows the best-fitting Gaussian to the distribution. The vertical dashed lines denote $\pm 4\sigma_D$, which we adopt as our variability threshold.

We find that 7 unresolved, matched sources in our survey area exceed the $4\sigma_D$ threshold, and thus exhibit variability on the $\sim10$-year baseline probed by comparing our observations to the FIRST\ catalog; their properties are given in Table~\ref{tab:var}.  The detected variable source density is in reasonable agreement with the results of \citet{deVries04}, who use a similar catalog and approach to find 1 variable source per square degree over 120.2 deg$^2$ of high-latitude sky on a 7-year baseline:  4/7 of our variable sources have a fractional variability below 50\% (col.\ 5 of Table~\ref{tab:var}), while \citet{deVries04} report $(73 \pm 4)\%$.

 We therefore conclude that on timescales of years, the variable source density and degree of discrete source variability in the Draco field is typical of that measured in other regions of the high-latitude sky. We investigate the impact of this variability on our search for an extended radio halo in Draco in \S\ref{halo:var}.

\begin{deluxetable}{ccccc}
\tablecaption{Variable Sources in the Draco Field \label{tab:var}}
\tablehead{ \colhead{$\alpha,\delta$} & \colhead{$F_F$}  & \multicolumn{2}{c}{$|{F_m-F_F}|$} & \colhead{$FR$} \\
            \colhead{(J2000)} & \colhead{(mJy)}   & \colhead{(mJy)} & \colhead{($\sigma_D$)} &  \\
            \colhead{(1)}     & \colhead{(2)}    & \colhead{(3)}   & \colhead{(4)}  & \colhead{(5)}           }
\startdata
$\mathrm{17^h05^m04^s.5}, \, 60\arcdeg03\arcmin56\arcsec$ & 3.76  & 1.81 & 4.53 & 1.48 \\
$\mathrm{17^h19^m37^s.3}, \, 58\arcdeg47\arcmin55\arcsec$ & 27.3 & 7.18 & 25.2 & 1.36 \\
$\mathrm{17^h27^m50^s.7}, \, 57\arcdeg51\arcmin13\arcsec$ & 2.03 & 1.40 & 5.43 & 1.69\\
$\mathrm{17^h08^m02^s.3}, \, 57\arcdeg44\arcmin03\arcsec$ & 4.32  & 1.94 & 7.22 & 1.81\\
$\mathrm{17^h06^m50^s.1}, \, 56\arcdeg56\arcmin15\arcsec$ & 4.33  & 2.08 & 5.81 & 1.92\\
$\mathrm{17^h25^m45^s.7}, \, 56\arcdeg28\arcmin18\arcsec$ & 7.24 & 1.22 & 4.15  & 1.20\\
$\mathrm{17^h06^m57^s.4}, \, 55\arcdeg54\arcmin42\arcsec$ & 3.54 & 1.20 & 4.08 & 1.34
\enddata
\tablecomments{Col.\ 1: FIRST\ source position. Col.\ 2: FIRST\ source peak flux. Cols.\ 3, 4: absolute (measured - FIRST) peak flux difference in mJy and in units of the net statistical uncertainty. The FIRST\ clean bias and median flux difference for the sample were subtracted. Col.\ 5: fractional variability, or the ratio of the brightest to the faintest flux ($FR > 1$).}
\end{deluxetable}

\section{Searching for Extended Emission}
\label{halo}

We wish to assess the likelihood that the subtracted Stokes I GBT maps in Fig.~\ref{fig:sub} contain emission that could constitute an extended dark matter annihilation halo. To quantify this, we compute the azimuthally averaged radial surface brightness profile in each GBT map relative to the stellar centroid of the dSph therein, as well as the statistical significance of features in the profiles (\S\ref{halo:prof}). We then establish that uncertainties due to variability in the discrete source populations subtracted from the maps have a negligible impact on the derived profiles (\S\ref{halo:var}). Finally, we quantify our ability to recover extended radio halo profiles given the baselining procedure adopted to calibrate the data (\S\ref{halo:bias}).
%, and use the results of this exercise in concert with our profile uncertainties to constrain the brightness of any putative radio halos associated with each targeted dSph (\S\ref{halo:limit}).

\subsection{Radial Profile Derivation}
\label{halo:prof}

We carry out the bulk of our analysis on radial profiles computed from the subtracted Stokes I maps in Fig.~\ref{fig:sub}. The reason for this is twofold. First, because the observations are dominated by mapping uncertainties (c.f.\ \S\ref{data:gbt}) it is difficult to characterize the noise in each pixel, which precludes computing reliable statistics directly from the maps. However, uncertainties on radial profile values can be robustly determined by jacknifing the data. Second, the CPU07 radio halo prediction is azimuthally symmetric (Fig.~\ref{fig:pred}), and as a result the comparison of this prediction to profiles computed from the maps is straightforward. 

We derive radial profiles $I(r)$ from the subtracted Stokes I maps by extracting the mean value of all points that fall in circular rings as a function of distance from the stellar centroid of each dSph (Table~\ref{tab:basic}). We note that the uncertainties on these centroids are much smaller than the resolution of our GBT maps, and varying the point about which the profile is computed within these uncertainties has no impact on the results.

We exploit the difference maps of each region to obtain realistic estimates of the uncertainties on each profile point $I_{i}$. We generate 5000 ``shifted" realizations of each difference map: for each realization, we shift the map (and corresponding weights) by a random number of pixels in RA and dec, with pixels shifted outside the map region wrapped to the other side of it. We then compute the radial profile for each realization: for perfect data, this radial profile should have $I(r)=0$ regardless of the structure present in Stokes I. For each difference map, the uncertainty on each radial profile value $I_i$ due to mapping errors is then the standard deviation of the points obtained from the 5000 realizations. We then adopt the uncertainties on the profile points derived from the difference maps in Fig.~\ref{fig:data}i-\ref{fig:data}l (that have the largest standard deviations for each field) as those on the profiles derived from the Stokes I maps. 
% and the uncertainties on the subtracted Stokes I maps for each region are given . The uncertainties on the subtracted Stokes I maps of each region are then given by:
%\begin{equation}
%\sigma_i = \max \left( \frac{\sigma^j_{mD,i}}{2} \right) \,\,\,,
%\label{eq:sig_prof}
%\end{equation}
%The uncertainties on the radial profiles for the split maps are then $\sigma^j_{mD,i}/\sqrt{2}$, where the factor of $\sqrt{2}$ arises because the difference map equals Split A - Split B and again $j$ is one of time, elevation or PA. Finally, 
%where $j$ corresponds to PA for Draco, UMaII and Coma, and to elevation for Will1 (c.f. Fig.~\ref{fig:data}i-\ref{fig:data}l). 

\begin{figure*}
\epsscale{1.15}
\plottwo{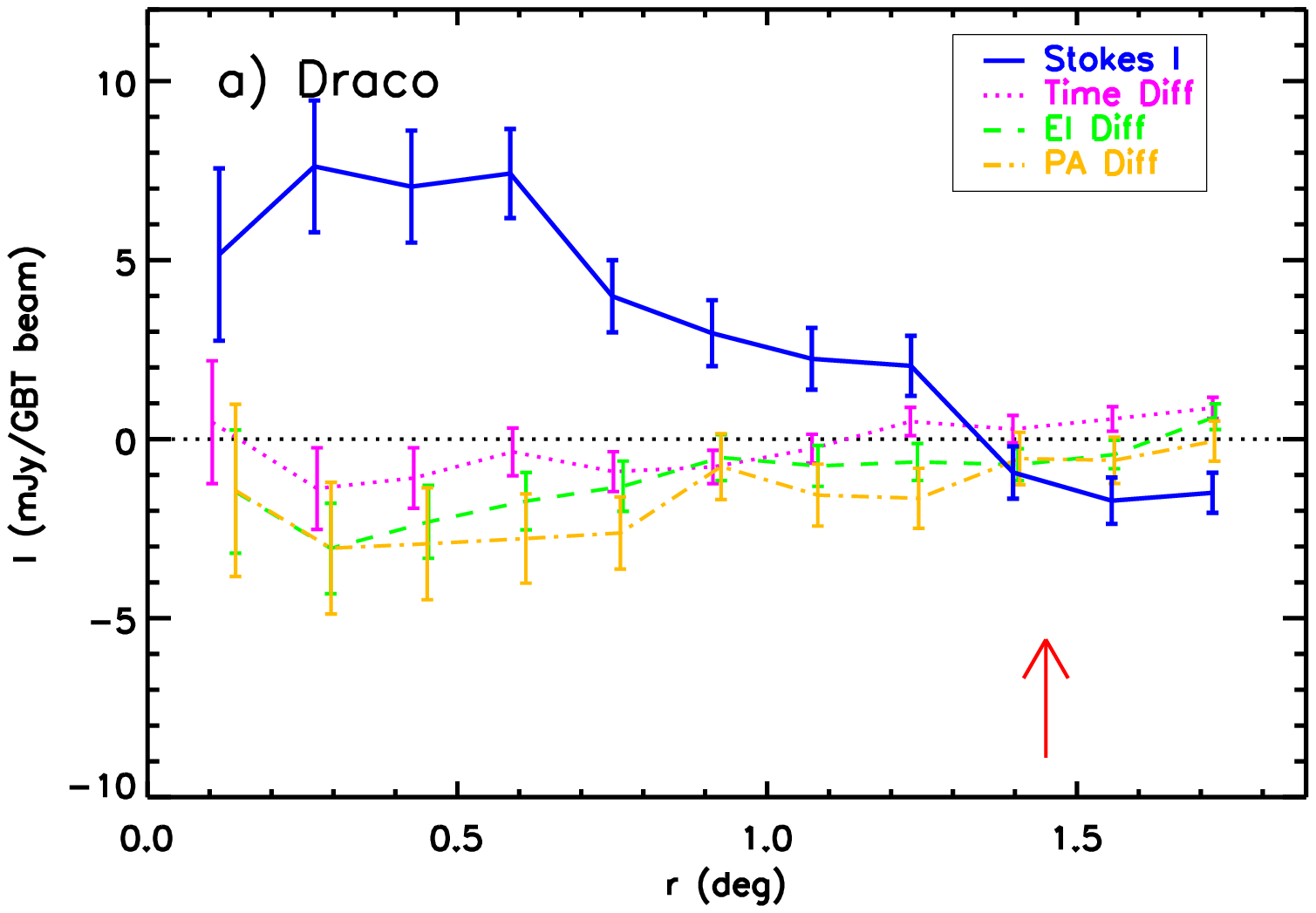}{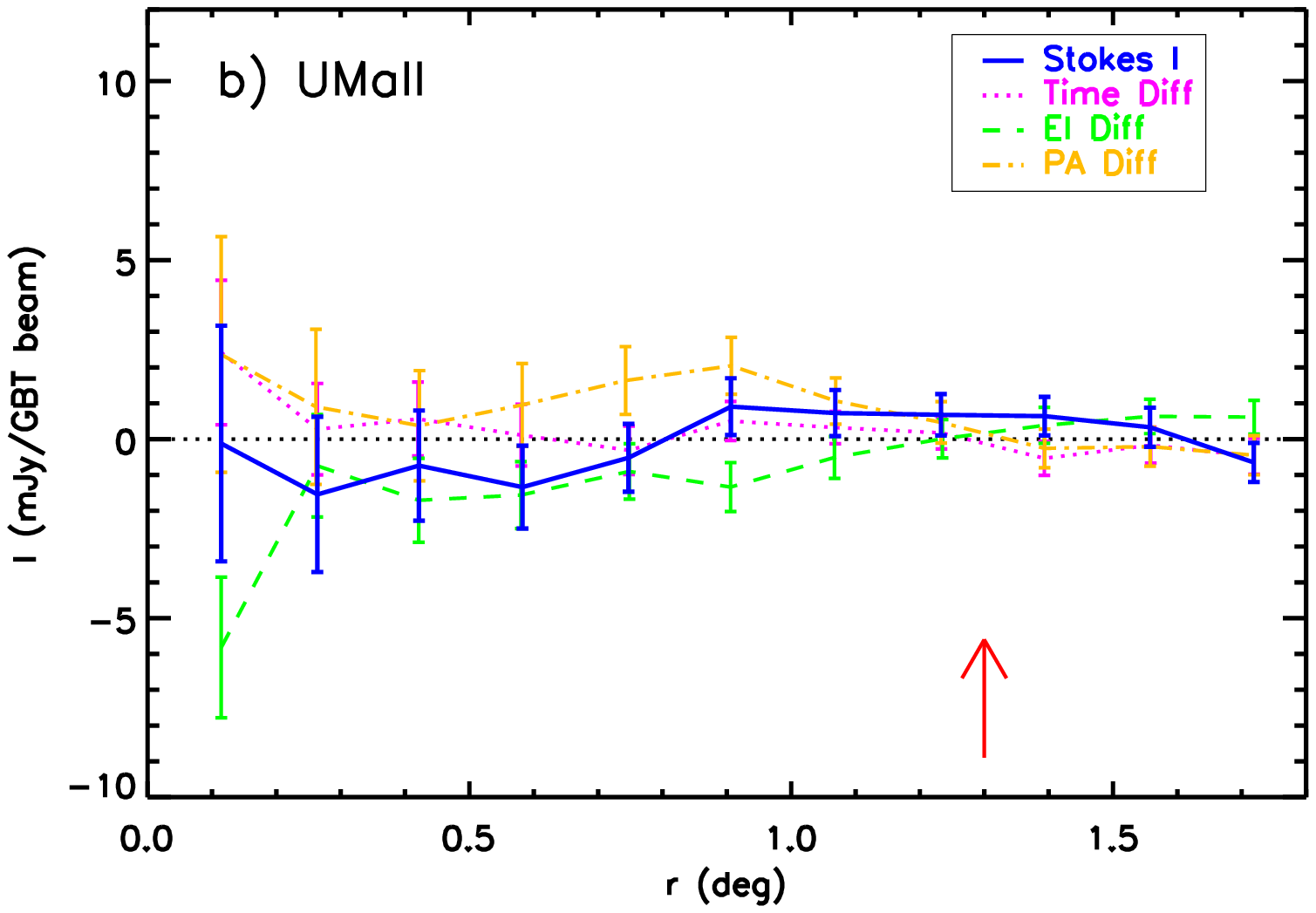}
\plottwo{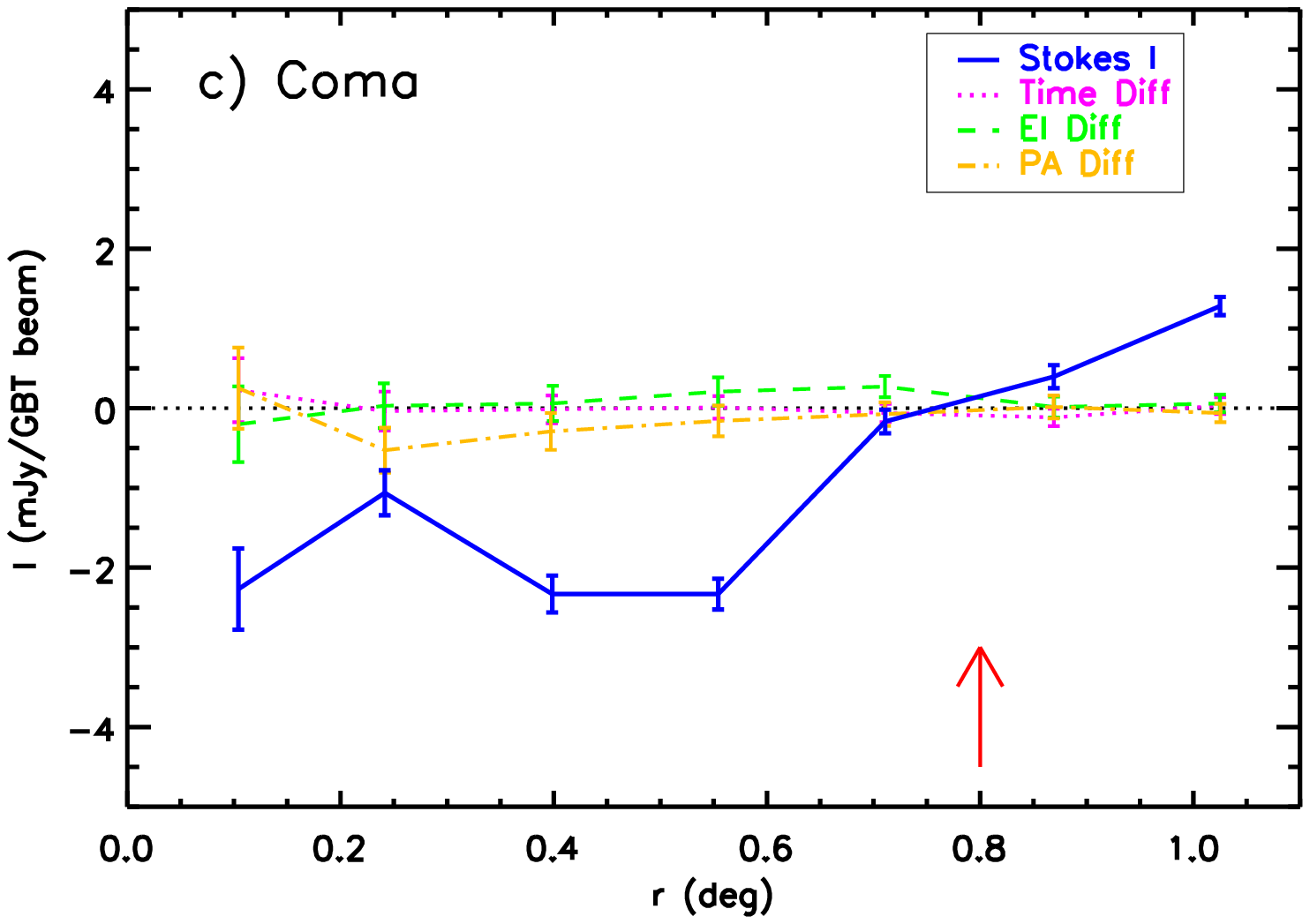}{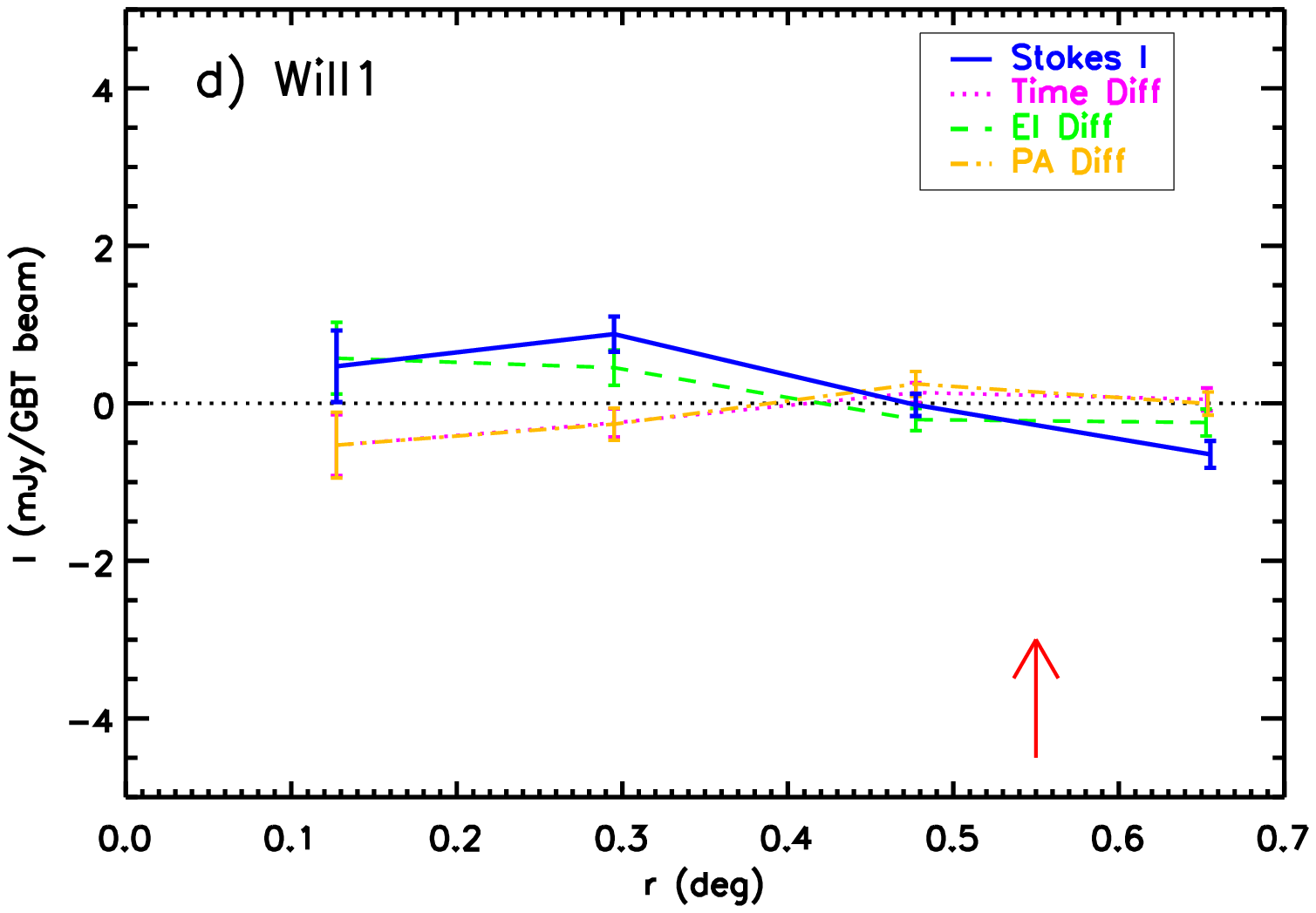}
\caption{Radial profiles measured from the subtracted Stokes I GBT maps in Fig.~\ref{fig:sub} (solid line) for a) Draco, b) UMaII, c) Coma and d) Will1. Note that the horizontal and vertical scales in the panels differ. In each panel, the dotted, dashed and dot-dashed lines show profiles derived from the time, elevation and PA difference maps, respectively. For perfectly mapped fields, these profiles would have $I(r)=0$. Points to the left of the vertical arrow were included in the $\chi^2_r$ computation of Table~\ref{tab:stats} and the comparisons to the halo models of CPU07 in \S\ref{discuss:cpu07}; see text for details.\\
(A color version of this figure is available in the online journal.)}
\label{fig:prof}
\end{figure*}

Fig.~\ref{fig:prof} shows radial profiles derived from the subtracted maps in Fig.~\ref{fig:sub} (solid line), as well as those derived from the time (dotted line), elevation (dashed line), and PA (dash-dotted line) difference maps for each region. Qualitatively, all three difference map profiles for each field seem consistent with $I (r) = 0$, as expected if the telescope response has been properly calibrated out of the data and our error analysis is reliable. The profiles derived from the subtracted maps in the UMaII and Will1 fields (Figs.~\ref{fig:prof}b~and~\ref{fig:prof}d) also seem to contain little structure.  Conversely, the Draco profile (Fig.~\ref{fig:prof}a) has $I(r)>0$ (relative to the subtracted mean sky brightness) at $r<1.3\arcdeg$, while that for Coma (Fig.~\ref{fig:prof}c) has $I(r)<0$ at $r<0.7\arcdeg$. 

We quantify the statistics of the radial profiles by computing the reduced $\chi^2$ statistic for the hypothesis that $I(r) = 0$. We take into account correlated mapping uncertainties by using the covariance matrix derived from the profiles of the 5000 difference map realizations. Specifically, for each difference map $j$, the $\chi^2_r$ statistic is given by: 
\begin{equation}
\chi^2_{r,Cj} = \frac{1}{N} \,\mathrm{I_j \, C_j^{-1} \,I_j^T} \,\,\,,
\end{equation}
where $\mathrm{I_j}$ is a vector containing $N$ radial profile points corresponding to that difference map, $\mathrm{C_j}$ is the ($N \times N$) covariance matrix computed from the profiles of its 5000 realizations, and $j$ is one of time, elevation or PA. 
%Note that if the diffference map profile uncertainties are uncorrelated (ie. $\mathrm{D_j}$ is diagonal), then this relation reduces to the standard $\chi^2 = \frac{1}{N} \sum_i(I_i/ \sigma)^2$. 
The value of $\chi^2_r$ for the Stokes I profile for each field  is then obtained from: 
\begin{equation}
\chi^2_r = \min \left[ \frac{1}{N} \,\mathrm{I \, C_{j}^{-1} \,I^T} \right ]\,\,\,,
\label{eq:chieq}
\end{equation}
where I is a vector containing $N$ radial profile points from one of the Stokes I maps in Fig.~\ref{fig:sub}.  In words, we use the difference map covariance matrix that produces the smallest $\chi^2_r$ to assess the significance of the features in each Stokes I profile. In practice, the $C_j$ used correspond to the difference maps shown in Fig.~\ref{fig:data}i-\ref{fig:data}l, whose pixel standard deviations are largest. 
%The of the difference map in Fig.~\ref{fig:data}i-\ref{fig:data}l to assess the significance of the emission in each field. 
%As expected, the matrices used correspond to the difference maps shown in Fig.~\ref{fig:data}i-\ref{fig:data}l, whose pixel standard deviations are largest. 
%Note that the factor of 4 in eq.~\ref{eq:chieq} has the same origin as that in eq.~(\ref{eq:sig_prof}). 

Table~\ref{tab:stats} shows $\chi^2_r$ obtained for the discrete-source subtracted Stokes I maps and the difference maps for each region. For each entry, the number in parentheses is the one-sided $p$-value of the chi-squared test for that entry and the number of degrees of freedom in col.~2. Because the number of pixels contributing to each profile point increases with $r$, the outermost profile points strongly influence the reported statistics. We are most interested in quantifying detection statistics near each dSph at the centre of the maps, and therefore compute $\chi^2_r$ for the profile points at smaller $r$ than the vertical red arrow in each panel of Fig.~\ref{fig:prof}; none of our conclusions change if all profile points in each field are included. 

Table~\ref{tab:stats} confirms that there is no statistically significant structure in any of the difference maps for any field, and therefore that our calibration is effective and that the difference maps provide a reasonable estimate of the mapping uncertainties for each field. We note that the statistics of the difference map radial profiles change little if correlated errors are ignored in the analysis: the profile points are largely uncorrelated. 
Among the radial profiles for the discrete-source subtracted maps of each region, both Draco and Coma show statistically significant structure at the $7.7\sigma$ and $43\sigma$ level, respectively.  As we discuss in \S\ref{discuss}, we attribute this structure to foregrounds. There is no statistically significant structure in the UMaII field, while that in the Will1 field is marginally significant at the $5.8\sigma$ level. 
%The significance of all features in the Stokes I map profiles increases if correlated uncertainties are ignored.

%\begin{figure}
%\epsscale{1.1}
%\plotone{FIGS/nosub.ps}
%\caption{Radial profiles measured from the unsubtracted GBT map in Fig.~\ref{fig:nosub}a (thick dash-dotted lines) and the discrete-source subtracted GBT map in Fig.~\ref{fig:sub} (thick dashed lines). 
%The mean flux is measured in circular rings about Draco's photometric centre, and the errorbars are the standard deviation on the mean. 
%The dotted lines near each thick curve show the profiles measured from the two (nearly) independent polarizations averaged to produce the final maps. The thick solid line is the derived upper limit (at 95\% confidence) on the flux of a putative radio halo in Draco (see \S\ref{halo:limit}).\\
%(A color version of this figure is available in the online journal.)}
%\label{fig:prof}
%\end{figure}
%\end{figure}

\begin{deluxetable*}{lccccc}
\tablecaption{Radial Profile $\chi_r^2$ Statistics \label{tab:stats}}
\tablehead{ \colhead{Field} & \colhead{DOF} & \colhead{Stokes I} & \colhead{Time Diff} & \colhead{El Diff} & \colhead{PA Diff} \\
                    \colhead{(1)}         & \colhead{(2)}       & \colhead{(3)}          & \colhead{(4)}       & \colhead{(5)}  & \colhead{(6)}  }
\startdata
Draco         & 9 & 7.7($<$0.0001) &  1.6(0.11) &  1.1(0.36) & 2.2(0.019)\\
UMaII           & 8 & 0.8(0.60)   &  2.1(0.032) &  0.5(0.86) & 1.0(0.43) \\
Coma         & 5 & 43($<$0.0001) &  1.1(0.36) &  0.1(0.99) & 1.1(0.36) \\
Will1   & 3 & 5.8(0.00078)   &  1.6(0.12) &  1.3(0.27) & 1.2(0.31) 
\enddata
\tablecomments{Col.~1: Field name. Col.~2: Number of degrees of freedom in $\chi^2_r$ computation. Cols.~3--6: Value of $\chi^2_r$ obtained for the Stokes I (col.~3), time difference (col.~4), elevation difference (col.~5) or PA difference (col.~6) profile. The number in parentheses is the one-sided $p-$value of the chi-squared test given $\chi_r^2$ and the degrees of freedom in col. 2.}
\end{deluxetable*}

\subsection{Impact of Discrete-Source Variability}
\label{halo:var}

The similarity between the unsubtracted Stokes I maps and the NVSS data for each region (see Fig.~\ref{fig:nosub} for Draco) and the absence of residuals at the locations of NVSS sources in our subtracted maps (Fig.~\ref{fig:sub}) suggest that discrete-source variability does not strongly impact the derived radial profiles for each region. Below, we quantify this statement using the variability analysis for the Draco field from \S\ref{data:vla}.

To estimate the importance of variability in the point-source population, we conservatively assume that the distribution of peak flux differences between our VLA map and the FIRST\ catalog (Fig.~\ref{fig:VLAvar}) for the Draco field stems entirely from intrinsic, low-level  variability in the discrete source population on timescales of years, and that this variability is present on the angular scales probed by the NVSS. In reality, one expects the variability of NVSS sources to be smaller than that of FIRST\ sources, because the extended radio emission resolved out by FIRST\ but retained by the NVSS typically varies on longer timescales than probed here. Simulations using the variability between our VLA observations and the corresponding FIRST\ map should therefore provide a conservative upper limit on the importance of this potential bias. 

We therefore generate a new realization of the subtracted Stokes I GBT map for each dSph, and add a residual at the location of each NVSS source therein whose amplitude is randomly drawn from the gaussian in Fig.~\ref{fig:VLAvar}. The probability of drawing a  $\geq 4\sigma$ outlier from this distribution is much lower than required to produce the 7 genuinely variable sources in the Draco field (see Table~\ref{tab:var}): to account for this, we also assign the peak flux differences of these variable sources to 7 randomly selected NVSS source locations in each realization. We add these 7 sources to all of our fields, even those  that are much smaller than the Draco field (Coma and Will1): this mimics the effect of a larger space density of variable sources, which we address below. The ensemble of residuals for each realization is convolved to the GBT resolution and added to the subtracted map for the corresponding field, and the radial profile is measured. We repeat this exercise 5000 times for each field to examine the variation in the measured profiles.

\begin{figure}
\epsscale{1.1}
\plotone{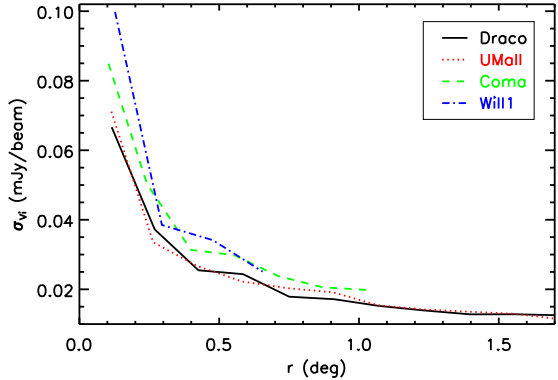}
\caption{Estimated uncertainty in the discrete-source subtracted Stokes I profiles in Fig.~\ref{fig:prof} due to variability in the discrete source population. The lines show the standard deviation of profile points derived from 5000 realizations  of the Draco (solid line), UMaII (dotted line), Coma (dashed line) and Will1 (dot-dashed line) fields obtained by adding residuals drawn from the Gaussian distribution in Fig.~\ref{fig:VLAvar} as well as the 7 outliers from Table~\ref{tab:var} randomly at the locations of NVSS sources. \\
(A color version of this figure is available in the online journal.)}
\label{fig:var}
\end{figure}

 Fig.~\ref{fig:var} shows the standard deviation $\sigma_{Vi}$ at each profile point computed from the 5000 realizations in each of the mapped fields. For all $r$ in all fields, these values are at least an order of magnitude smaller than the uncertainties on the radial profile points (c.f.\ Fig.~\ref{fig:prof}): $\sigma_{Vi} <<  \sigma_i$.   Even in the Coma and Will1 fields, where the space density of variable sources is assumed to be $\sim 2.6$ and $\sim 7$ times larger than in Draco by virtue of the addition of all 7 sources in Table~\ref{tab:var}, the resulting $\sigma_{Vi}$ increases by at most $30\%$. It is clear that variability in the discrete source population contributes negligibly to the uncertainty in the measured radial profiles in Fig.~\ref{fig:prof}.

 The space density of bright ($F \gtrsim 100\,$mJy) transient or strongly variable radio sources is estimated to be $\lesssim 0.004\,$deg$^2$ (e.g.\ \citealt{gregorini86}; \citealt{croft10}; but see \citealt{matsumura09}), and they are therefore quite rare in degree-sized fields like those considered here. Nonetheless, it is possible for one such event to escape detection in our variability study (which only probed one $\sim10$-year baseline) and produce a large residual in the discrete-source subtracted map. This scenario seems unlikely for our fields, since the largest residuals therein does not coincide with the location of NVSS sources. Nonetheless, we examined the impact of adding an unresolved residual, with an amplitude equal to the largest in the Draco field, at random locations in the subtracted Stokes I maps. Unless one is unfortunate enough to have this residual fall within $\sim2$ GBT beams of the profile center, this single large residual does not strongly alter the measured radial profiles. In practice, it is straightforward to mask out large residuals when the profile is computed, and we therefore consider the effect of this uncertainty to be negligible.

\subsection{Sensitivity to Extended Emission}
\label{halo:bias}

 A key step in processing single-dish radio continuum observations is the removal of baseline drifts in the data, which we parametrize with $p_i$ in equation~(\ref{eq:red1}). Because these drifts have a characteristic scale on the order of the map size, an important side-effect of this step is that at least some large-scale emission is filtered out.  This is a well-studied effect of the subtraction of polynomial or other time-domain functions when making maps from time series scans \citep[see also][]{dicker09,aguirre11}.  We find that fitting a second-order polynomial to the data time-series is too aggressive, and filters out most of the annihilation halo flux predicted by CPU07.  We have therefore implemented a  linear baselining procedure, and evaluate its filtering effect here.
 
% For example, we find that fitting a standard, second-order polynomial to the baseline filters out most of the annihilation halo flux predicted by CPU07 \citep[see also][]{dicker09,aguirre11}.  We have therefore carefully evaluated the linear baselining procedure described in \S\ref{data:gbt} to determine its filtering effect.  Here, we quantify our sensitivity to the radio halo profiles that we are trying to detect.
 
 It is clear that the sensitivity of the final maps to a given emission feature is a function of both its characteristic scale and morphology: to use the extremes as an example, our maps have full sensitivity to discrete sources but are blind to constant emission across the field, as the latter would be completely filtered out when the data are baselined. The most reliable way to quantify our sensitivity to extended halos is therefore to inject them directly into the time-ordered data and to attempt to recover them post-processing. We carry out this procedure as follows: we inject each of the profiles in Fig.~\ref{fig:pred}, scaled radially to the $r_h$ of the dSph in each field (Table~\ref{tab:basic}), directly into the time-ordered data $d_i$ to produce a ``data+halo" input $d_i + h_i$. We then process $d_i + h_i$ in the same manner as for $d_i$ in \S\ref{data:gbt}. The maps produced from $d_i$ are subtracted from those produced from $d_i+h_i$: to zeroth order, this eliminates thermal and mapping noise as well as the input $d_i$ to produce an image of $h_i$ after processing. Finally, we compute radial profiles from the processed $h_i$ maps in the same manner as for the $d_i$ for each field.
 
 We verified that the reconstructed surface brightness of the processed halo images is linear with input halo surface brightness to within 2\% for $\mathrm{5\,mJy/beam} < I < \mathrm{100\,mJy/beam}$, as expected given the success of our discrete source subtraction [c.f.\ equation~(\ref{eq:red1})]. We therefore produced processed images for a single ``set \#1" and ``set \#2" halo predicted by CPU07 for the dSph in each field.
% , and we vary the intensity of these processed profiles to fit the data and place constraints on the extended radio emission in the targeted dSphs.
   
\begin{figure}
\epsscale{1.1}
\plotone{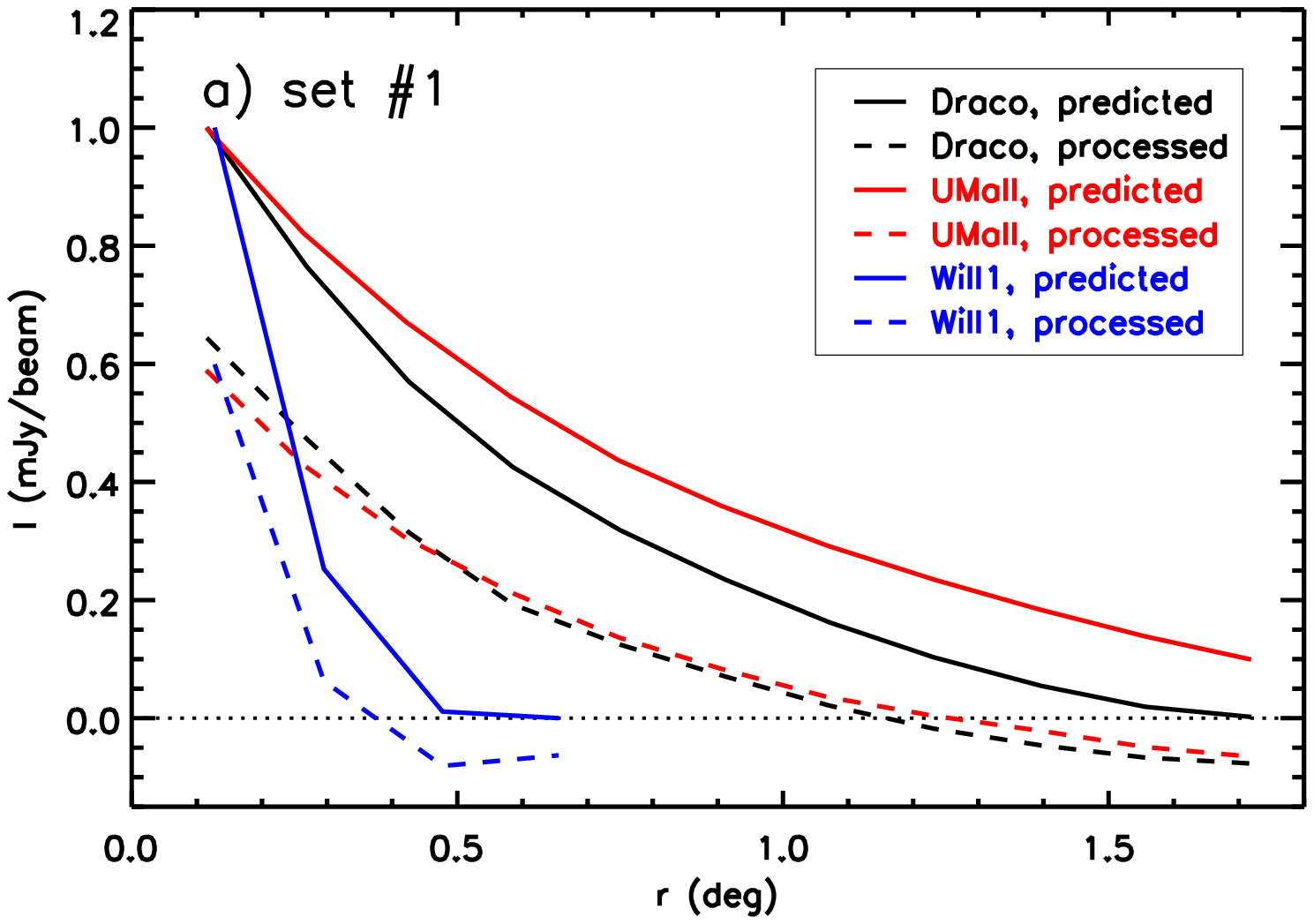}
\plotone{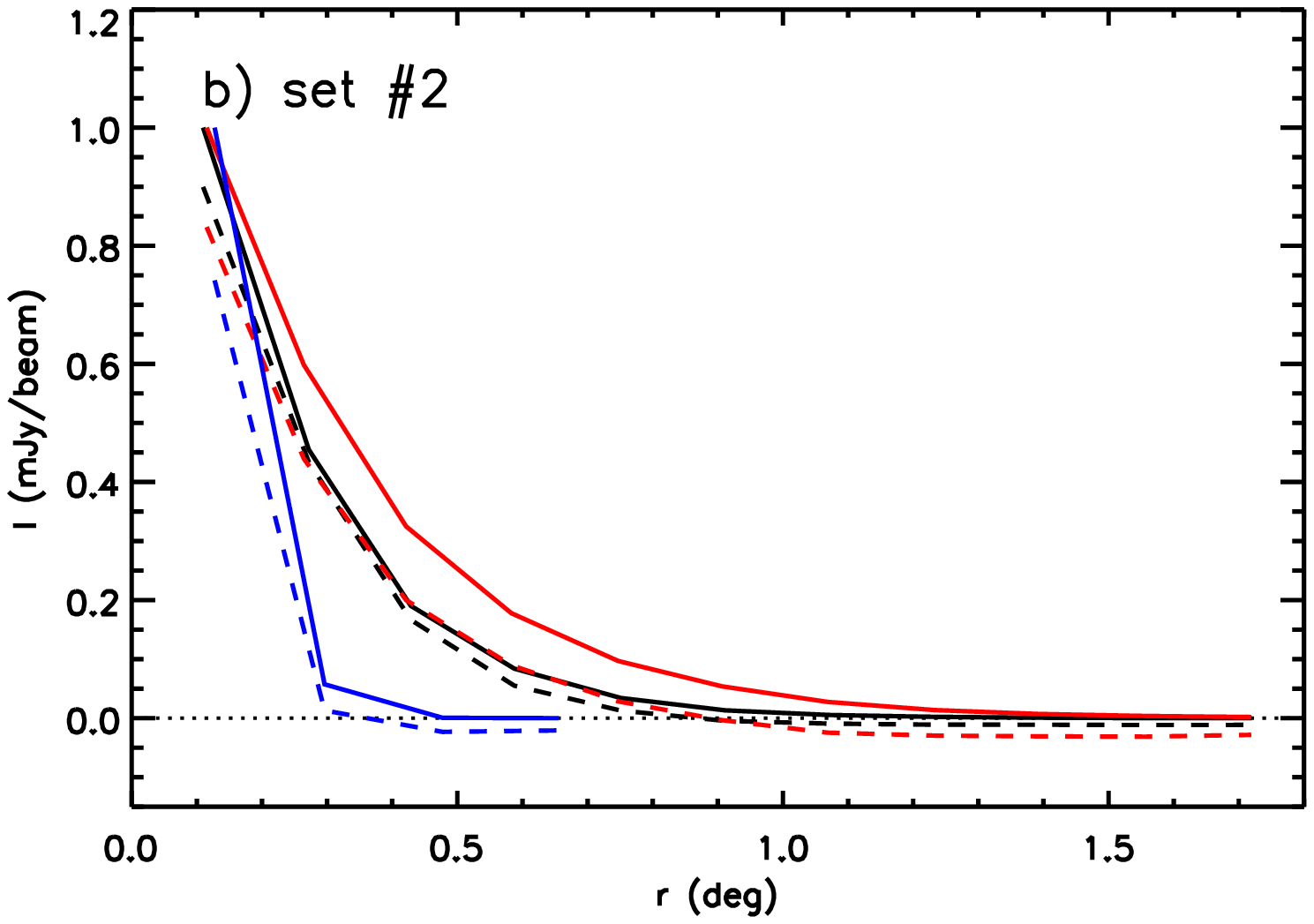}
\caption{Impact of baselining procedure on CPU07 predictions for a) ``set \#1" and b) ``set \#2". In each panel, the solid black, dark gray (red in the online version) and light gray (blue in the online version) lines show the predicted profiles from Fig.~\ref{fig:pred}. The profiles are scaled radially to the $r_h$ of the dSph of interest, evaluated at the measured profile radii of Fig.~\ref{fig:prof} and normalized so that the innermost profile point has $I_1=1\,$mJy/beam. The dashed lines of the same color show the profiles computed after processing the predictions through our pipeline, as described in \S\ref{halo:bias}. The processed profiles are both filtered and exhibit curvature bias, the extent of which depends on the angular size of the predicted halo. \\
(A color version of this figure is available in the online version of the Journal.)}
\label{fig:bias}
%\end{figure}
\end{figure}

   Fig.~\ref{fig:bias} illustrates the impact of our processing pipeline on different input halos. The solid lines therein show ``set \#1" (Fig.~\ref{fig:bias}a) and ``set \#2" (Fig.~\ref{fig:bias}b) profiles evaluated at the measured profile radii for each of Draco, UMaII and Will1. For convenience, we normalize the innermost profile point to $I_1=1\,$mJy/beam. The dashed lines in the same color show the corresponding radial profile computed from the processed halo maps. As expected, our sensitivity to a given profile depends on its radial extent relative to the map size: for example, our processing pipeline recovers only $\sim60\%$ of the peak surface brightness for ``set \#1" in Draco, but almost all of the predicted ``set \#2" halo flux.  Because a larger fraction of the ``set \#2" flux is recovered, we are correspondingly more sensitive to \sigv\ computed from this profile than for ``set \#1": our maps will therefore yield a range of constraints on \sigv\ by virtue of the propagation parameters adopted. 
   
   Note that in Fig.~\ref{fig:bias}, the processed ``set \#1" halo profiles are negative at large $r$: this is another artifact of the baseline subtraction. By removing a mean and slope from the maps, the linear baselining routine effectively imparts a ``curvature bias" relative to the map center.  Note, however, that this curvature bias does not imply a positive bias in large scale flux. The curvature bias could explain why the non-zero radial profiles for Draco and Coma, which we attribute to foregrounds below, appear to peak roughly at the map center. We explore mapping strategies to mitigate this and the filtering effect described above in \S\ref{discuss:fmap}.
   
    Throughout this paper, we compare all data products to the processed halo profiles (c.f.\ Figs.~\ref{fig:otherlim}~and~\ref{fig:dracolim}) to account for any filtering or curvature bias due to our baselining procedure when placing constraints on the extended emission from the targeted dSphs.
      
 %%%%%%%%%%%%%%%
 %%%%%%%%%%%%%%%  
\section{Discussion: Implications for WIMPs and Future Prospects}
\label{discuss}

 With measured radial profiles (\S\ref{halo:prof}), the impact of variability explored (\S\ref{halo:var}) and an assessment of our sensitivity to extended radio emission in hand (\S\ref{halo:bias}), we turn now to the implications of the subtracted Stokes I maps for WIMPs. Specifically, we interpret our derived radial profiles in the context of the CPU07 predictions  in \S\ref{discuss:cpu07}, deriving upper limits on \sigv\ from the UMaII and Will1 fields (\S\ref{discuss:cpu07:limits}) and excluding WIMP annihilations as the source of the features in the Draco and Coma fields (\S\ref{discuss:cpu07:draco}). We devote \S\ref{discuss:foregrounds} to a discussion of the limitations on our work imposed by foregrounds, and address the plausibility of invoking large-scale magnetic fields in dSphs in \S\ref{discuss:bfields}. In \S\ref{discuss:future}, we exploit the lessons learned in this study to suggest three avenues by which our constraints on \sigv\ can be improved and refined.  
 
\subsection{Comparison to CPU07 Models}
\label{discuss:cpu07} 
 
In order to provide an order-of-magnitude estimate of the \sigv\ that are probed by our maps, we compare the derived radial profiles in Fig.~\ref{fig:prof} to the fiducial CPU07 predictions shown in Fig.~\ref{fig:pred} (see \S\ref{intro}). In \S\ref{discuss:cpu07:limits}, we derive upper limits on \sigv\ from the UMaII and Will1 profiles, and compare these limits to those obtained from the two-year Fermi-LAT data. In \S\ref{discuss:cpu07:draco}, we argue that the emission in the Draco field most likely stems from foregrounds rather than WIMP annihilations.

\subsubsection{Upper Limits on \sigv\ from UMaII and Will1}
\label{discuss:cpu07:limits}

The radial profiles in Fig.~\ref{fig:prof} and statistics of Table~\ref{tab:stats} show that there is no significant structure in the UMaII field, while that in Will1 is only marginally significant. These fields are thus devoid of both extended radio halos as well as significant astrophysical foregrounds (a fortuitous cancellation of these two components is very unlikely, as borne out by our analysis in \S\ref{discuss:foregrounds:bayesian}). We therefore derive upper limits on the strengths of the radio halos in these dSphs allowed by the data: specifically, we assume that the emission from the dSphs in these fields is consistent with zero, and we use the errorbars on the profile points to derive upper limits on \sigv.  We restrict ourselves to comparisons between our data and the fiducial CPU07 models shown in Fig.~\ref{fig:pred} to produce order-of-magnitude constraints on \sigv.

%These dSphs thus lie in regions with low astrophysical foregrounds even at our sensitivity and resolution, and exhibit no evidence for extended radio emission. Assuming that the emission from the dSphs in these fields is consistent with zero, we use the errorbars on the profile points in each field to place upper limits on the strengths of the radio halos allowed by the data. 
  
 Fig.~\ref{fig:otherlim} shows the result of this exercise.  The hatched regions in each panel show the uncertainties on our radial profile measurements, and the solid lines in each panel show the CPU07 ``set \#1" and ``set \#2" predictions -- processed as described in \S\ref{halo:bias} -- normalized to be consistent with the profile uncertainties at 95\% confidence. Because the outermost profile points pull strongly on the fit but are of little astrophysical interest, we include only points to the left of the vertical arrow each panel in the fit, as in the $\chi^2_r$ computations of Table~\ref{tab:stats}. The corresponding upper limits on \sigv\ are given on the right-hand side of the panel. 
  
 Fig.~\ref{fig:otherlim} illustrates that our upper limits on \sigv\ from UMaII and Will1 differ by at least an order of magnitude between the CPU07 ``set \#1" and ``set \#2". This arises because the latter is more centrally concentrated than the former, which has two effects. First, for a given total flux, a centrally concentrated profile is easier to detect than one distributed over more pixels. Second, our baselining algorithm filters out less flux for a centrally concentrated profile relative to a more extended one (see \S\ref{halo:bias}). 
 %This latter effect is particularly important for UMaII (Fig.~\ref{fig:otherlim}a), which explains why the constraint from ``set \# 1" is weaker by $0.3\,$dex relative to that from Will1. 
 
 To compare our upper limits on \sigv\ with those from Fermi-LAT we first discuss the validity of the CPU07 diffusion models, which have the form $D(E) \propto D_0E^\gamma$ (see \S\ref{intro}). The values of $D_0$ adopted in CPU07's  ``set \#1" and ``set \#2" roughly span the $\sim3\sigma$ range allowed by cosmic-ray flux models of the Milky Way \citep{maurin01,donato04}.   Scaling arguments from diffusion models of galaxy clusters suggest that $D_0$ may be an order of magnitude smaller in dSphs than in the Milky Way, implying that ``set \#2" is most appropriate for dSphs \citep{jeltema08}.  Although the CPU07 values of $\gamma$ are significantly smaller than $0.46 \lesssim \gamma \lesssim 0.85$ allowed for the Milky Way \citep{maurin01}, this parameter does not strongly affect the predicted annihilation flux at $\nu = 1.4\,$GHz \citep{jeltema08}.  Recall that we scale the CPU07 profiles radially to account for different $r_h$ of the dSphs (see Fig.~\ref{fig:pred}), which also introduces uncertainty. Finally, we discuss plausible magnetic field values for dSphs in \S\ref{discuss:bfields}, and argue that $B=1\,\mu$G is a reasonable assumption. We therefore find the CPU07 ``set \#2" model to be plausible for dSphs, albeit with considerable uncertainty: the predicted annihilation flux is likely valid to within an order of magnitude.
% Recall as well that we scale the CPU07 models radially to account for the larger angular size of UMaII relative to Draco, which itself introduces an uncertainty
  
   Fig.~\ref{fig:otherlim} shows that comparing to the CPU07 models in Fig.~\ref{fig:pred}, non-detections in the UMaII and Will1 fields imply and $\sigv \lesssim 10^{-25}\,\mathrm{cm^3\,s^{-1}}$ for ``set \#2" at 95\% confidence. For an identical decay channel, WIMP mass and statistical confidence, the two-year Fermi-LAT constraint for UMaII is $\sigv < 2.17\times10^{-25}\,\mathrm{cm^3\,s^{-1}}$ \citep{ackermann11}; that for Will1 is not computed (c.f.\ \citealt{abdo10}). Note that UMaII is the dSph for which the Fermi-LAT constraint is the tightest among the 10 dSphs in that study. Our upper limits on \sigv\ for the most plausible ``set \#2" diffusion parameters is therefore commensurate with those from the two-year Fermi-LAT data for individual dSphs.  
   
   We caution that a direct comparison between our limits and those from \citet{ackermann11} is muddled by several factors. Contrary to that study, we do not include J-value uncertainties in our upper limits. In addition, while random uncertainties due to foregrounds are encapsulated in the errorbars of Fig.~\ref{fig:prof}, systematic uncertainties are not (see \S\ref{discuss:foregrounds}). Finally, we do not include (the considerable) uncertainties in the CPU07 diffusion models in our upper limits.  All of these factors will weaken our constraints relative to those from \citet{ackermann11}. On the other hand, the J-values for UMaII and Will1 are $\sim0.5-0.8\,$dex larger than that implicit in the CPU07 models \citep{strigari08,ackermann11}: in this sense, our upper limits are conservative. 
   
   A suite of detailed, CPU07-like predictions tailored to UMaII and Will1 that will enable detailed comparisons between our limits on \sigv\ and those from \citet{ackermann11} is under construction (A. Natarajan et al, in preparation), but beyond the scope of the present work. Nonetheless, the comparisons of Fig.~\ref{fig:otherlim} indicate that deep radio observations of dSphs have the potential to provide interesting constraints on particle dark matter properties; as such, they are highly complementary to searches at $\gamma$-ray energies.

%   where the confidence interval and decay channel adopted are identical to the CPU07 models;
 %For ``set \#2", our upper limit on \sigv\ is commensurate with the two-year Fermi-LAT  limits for individual dSphs \citep{ackermann11,geringer-sameth11}. 
% , and within $\sim 0.5\,$dex of the thermal prediction $\log \sigv^{th} = -25.5$ (see \S\ref{intro}).
% We caution that a direct comparison between the Fermi-LAT \sigv\ and our limits is muddled by significant differences between the studies: most importantly, our limits do not include the uncert
 
 %muddled by differences in halo properties and WIMP models adopted, as well as uncertainties in the magnetic field and charged particle propagation parameters inherent in our constraint.  

 \subsubsection{Emission in the Draco Field}
 \label{discuss:cpu07:draco}
    
  Fig.~\ref{fig:prof} and Table~\ref{tab:stats} illustrate that $I(r) \neq 0$ in with high significance in the Draco and Coma fields. The depression near the Coma field center almost certainly stems from foregrounds, since an annihilation halo signal should be positive.  However, the radial profile for Draco shows a statistically significant emission feature with the same basic morphology as that predicted by CPU07.  
 %However, it is  premature to associate this emission with dark matter annihilation. 
 However, two arguments suggest that this emission also stems from foregrounds instead of dark matter annihilation, and therefore that both the Draco and Coma fields are contaminated by foregrounds.

 First, the subtracted Stokes I map in Fig. 5a exhibits a horizontal
band of emission passing through the center of Draco, which is more
suggestive of foregrounds than the expected symmetric emission centered
on the dSph. Indeed, the amplitude of the emission seen in the Draco field is within a factor of two of the depression in the Coma field center, suggesting that the features in both maps stem from astrophysical foregrounds on similar angular scales. Recall that the correlation of these features with the map center stems at least in part from the curvature bias imparted by our baselining procedure (\S\ref{halo:bias}).
 
  Second, there is tension between the \sigv\ implied by the Draco profile if foregrounds are negligible and the upper limits on this parameter from Fermi-LAT as well as from our UMaII and Will1 fields. Fig.~\ref{fig:dracolim} shows the best-fitting processed ``set \#1" and ``set \#2" predictions to the Draco profile points and the implied \sigv. It is clear that if one associates the majority of the emission in the Draco field with an extended halo, the \sigv\ required to produce the emission exceeds the Fermi-LAT two-year limits \citep{ackermann11,geringer-sameth11} and our  UMaII and Will1 limits (Fig.~\ref{fig:otherlim}). We therefore conclude that the Draco field is likely contaminated by foreground emission. We devote the next section to a thorough discussion of the limitations on our analysis due to foregrounds, and find that foregrounds cannot be reliably subtracted from the data in-hand; we therefore do not attempt to constrain \sigv\ from either the Draco or Coma fields. 

\subsection{The Impact of Foregrounds}
\label{discuss:foregrounds}

As explained in \S\ref{discuss:cpu07}, our analysis of the derived radial profiles in the context of WIMPs is foreground-limited: we derive upper limits on \sigv\ from the UMaII and Will1 fields because they appear nearly foreground-free (\S\ref{discuss:cpu07:limits}), while we do not constrain \sigv\ from the Draco and Coma fields because they are contaminated by foregrounds (\S\ref{discuss:cpu07:draco}). We expland on the limitations due to foregrounds in this section. We discuss avenues for mitigating foregrounds in radio observations in \S\ref{discuss:foregrounds:general}, and argue that none can be reliably applied to the data in-hand. We therefore invert the problem in our maps to produce high-resolution, high-sensitivity measures of Galactic foreground emission at $1.4\,$GHz. In \S\ref{discuss:foregrounds:bayesian}, we attempt to simultaneously disentangle annihilation halo and foreground signals using a joint Bayesian analysis of all four mapped fields: the results from this exercise validate our approach in \S\ref{discuss:cpu07:limits}.

\subsubsection{Mitigating Foregrounds}
\label{discuss:foregrounds:general}
 
A potential avenue for mitigating Galactic foregrounds -- which predominantly stem from synchrotron emission -- is to explore the spectral dependence of the features in the subtracted maps of Fig.~\ref{fig:sub}. The spectral index of the annihilation signature depends on the adopted propagation parameters (c.f.\ fig.\ 12 of CPU07), but it generally resembles $\alpha \sim -0.7$ expected for Galactic synchrotron emission for $1\,\mathrm{GHz} \lesssim \nu \lesssim 10 \mathrm{GHz}$, and steepens to $\alpha \lesssim -1$ for $10\,\mathrm{GHz} \lesssim \nu \lesssim 100 \mathrm{GHz}$. So in principle, deep $\nu > 10\,$GHz radio observations would enable spectral index analyses that may help to identify foregrounds and subtract them from the maps. In practice, however, mapping degree-scale fields at such high radio frequencies to the required sensitivity is prohibitively expensive. 
% will not be feasible until the advent of phased-array feeds, and even then the uncertainties in the determined indices are unlikely to afford an unambiguous determination of the foreground \red{(ref here)}.

We did attempt to examine extant survey data for the fields in Fig.~\ref{fig:sub} at lower frequencies from the compilation of diffuse Galactic radio emission by \citet{costa08}. Their predicted all-sky map at $\nu=1.4\,$GHz is largely an extrapolation of the \citet{haslam82} $\nu=408\,$MHz survey data, which has an angular resolution of $1\arcdeg$ and an average zero-level uncertainty of $3\,$K.  Processing these maps through our pipeline as described in \S\ref{halo:bias}, we find fluctuations on the order of $10\,$mJy/beam. However, scaling the \citet{haslam82} zero-point uncertainty to $\nu=1.4\,$GHz assuming synchrotron foregrounds, these fluctuations lie well within the noise. There are therefore no extant foreground maps that have either the sensitivity or resolution to compare to our subtracted maps. 

Another possibility for mitigating foregrounds is to use the Galactic \hi\ spectral line signals that we collected simultaneously with the continuum measurements presented here (see \S\ref{data:gbt}; J. Aguirre et al.\ 2013, in preparation). \hi\ is highly correlated with dust,  which motivates the use of GBT-obtained \hi\ maps to clean maps of Galactic foreground dust emission \citep{planckCIB}.  There is a correlation between the synchrotron and dust emission for entire galaxies which persists on very large scales, but the correlation on small scales is expected to be tight only for the highest radio frequencies \citep{bennett03}.  Thus we do not expect that the \hi\ emission will be a complete representation of foregrounds in our maps.
%We note that statistical foreground tracers such as \ion{H}{1} emission \citep{planckCIB} are not reliable for the relatively small sample and map sizes presented here.

We can therefore turn the problem around and assume that there is no extended radio emission from any of the targeted dSphs; then, the maps in Fig.~\ref{fig:sub} provide a constraint on high-latitude Galactic foregrounds at unprecedented sensitivity and resolution. In particular, $\sigma_{ast}$ in Table~\ref{tab:noise} is a measure of Galactic foregrounds at 10\arcmin\ resolution and $35\arcdeg \leq b \leq 84\arcdeg$. We find $1.8\,\mathrm{mJy/beam} \leq \sigma_{ast} \leq 5.7\,\mathrm{mJy/beam}$ for our fields with the largest $\sigma_{ast}$ at the lowest $b$ as expected. However, the correlation between $\sigma_{ast}$ and $b$ is not perfect, likely due to the small number of fields mapped. Nonetheless, our measurements represent an improvement of a factor of $\sim 50$ in sensitivity and $\sim6$ in angular resolution over current foreground measurements at $1.4\,$GHz, such as that computed by \citet{costa08}. 

%\red{Can we get a handle on foregrounds from our HI, james? Anything else to add?}

\subsubsection{A Bayesian Halo and Foreground Estimate}
\label{discuss:foregrounds:bayesian}

The Draco and Coma fields were discarded from the analysis in
\S\ref{discuss:cpu07} under the hypothesis that $I(r) \neq 0$ therein
stemmed from foreground contamination rather than extended radio
emission. Although the foreground argument seems to be the most
plausible interpretation of the available data, it raises the
uncomfortable possibility that $I(r) \simeq 0$ in the UMaII and Will1
fields results from a fortuitous cancellation of a positive halo
signal and negative foreground fluctuations: in this case our limits
on \sigv\ have little meaning. We regard this scenario as unlikely but
attempt here to simultaneously analyze all four mapped fields --
allowing for a contribution to $I(r)$ from both a CPU07 annihilation
halo and a foreground contribution -- in order to further validate the
approach in \S\ref{discuss:cpu07}.

We adopt a Bayesian approach using the best-fitting, processed CPU07 predictions to each Stokes I profile (as in Fig.~\ref{fig:dracolim} for Draco) and their uncertainties.  We assume that the best-fitting intensities are drawn from a normal distribution with the
following components: i) a positive definite annihilation halo signal, with a common normalization for all dSphs, ii) an RMS foreground
signal which can be either positive or negative, and iii)
our estimated zero-mean measurement noise. For the foreground signal
%,which is expected to be dominated by Galactic synchrotron radiation,
we consider two cases: a) a foreground with constant RMS power across all fields, and b) a foreground with a $1/\sin(b)$ dependence.  We
then find the most likely values of the halo normalization and the RMS
foreground contribution given the best-fitting processed ``set \#1''
and ``set \#2'' profiles to the measured Stokes I profiles in
Fig.~\ref{fig:prof} for all four fields jointly.

We find that the analysis strongly rules out the zero foreground case:
it is not possible to interpret the Stokes I profiles from all four
fields in the context of the CPU07 models without invoking a
foreground contribution. This is not unexpected given $I(r)<0$ in the
Coma field (Fig.~\ref{fig:prof}c), but the analysis nonetheless lends some quantitative
support to the approach adopted in \S\ref{discuss:cpu07}. We find that
an RMS foreground amplitude of $4 -60\,$mJy/beam is allowed by the
data at 95\% confidence, with a most likely value of $6 -
9\,$mJy/beam. This estimate is consistent with $1.8\,\mathrm{mJy/beam} \leq \sigma_{ast} \leq 5.7\,\mathrm{mJy/beam}$ obtained in \S\ref{data:gbt}, though with considerably
larger uncertainty because an extended halo is simultaneously modeled and because the Bayesian estimate assumes only one foreground fluctuation in each map (on the scale of the annihilation halo signal), whereas $\sigma_{ast}$ is computed over several independent beams.  
The most likely value of the halo
normalization peaks near zero for both ``set \#1'' and ``set \#2'',
justifying our computation of upper limits on \sigv\ from the UMaII
and Will1 profiles. However, including a foreground contribution
weakens the upper limits on \sigv\ by $\sim0.5\,$dex relative to those
derived in \S\ref{discuss:cpu07} for both foreground models considered. 
%This analysis sets $95\%$ upper
%limits on the normalization of the nominal CPU07 halo model of $0.64$
%(Set 1) and $0.01$ (Set 2). Both of these correspond to peak halo
%intensities of $25 -- 30 \, {\rm mJy/bm}$ for the GBT at $1.4 \, {\rm
%GHz}$. {\bf translate into <sigma v> - maybe we should also put
%multiplier limits in the other analysis? i think this is about 2-3 x
%weaker than the other analysis} 
%Results are not sensitive to the
%foreground model assumed for the two cases considered.

While the results of our Bayesian analysis are encouraging, we find
that our sample of four dSphs is too small to meaningfully distinguish foregrounds
from the signal of interest in a simultaneous fit: our approach in \S\ref{discuss:cpu07} of considering only fields that are unlikely to be contaminated by foregrounds is therefore preferable for the present study. Moreover, we have not considered the potential spatial dependence of the foreground signal within the maps, which is poorly constrained at their sensitivity and resolution (\S\ref{discuss:foregrounds:general}). Combined with the shortcomings of adopting a single halo normalization for all four fields, we find that a comprehensive Bayesian analysis is beyond the scope of this paper, but is likely to be both feasible and preferable as both larger samples and better models become available (see \S\ref{discuss:future}).

\begin{figure}
\epsscale{1.1}
\plotone{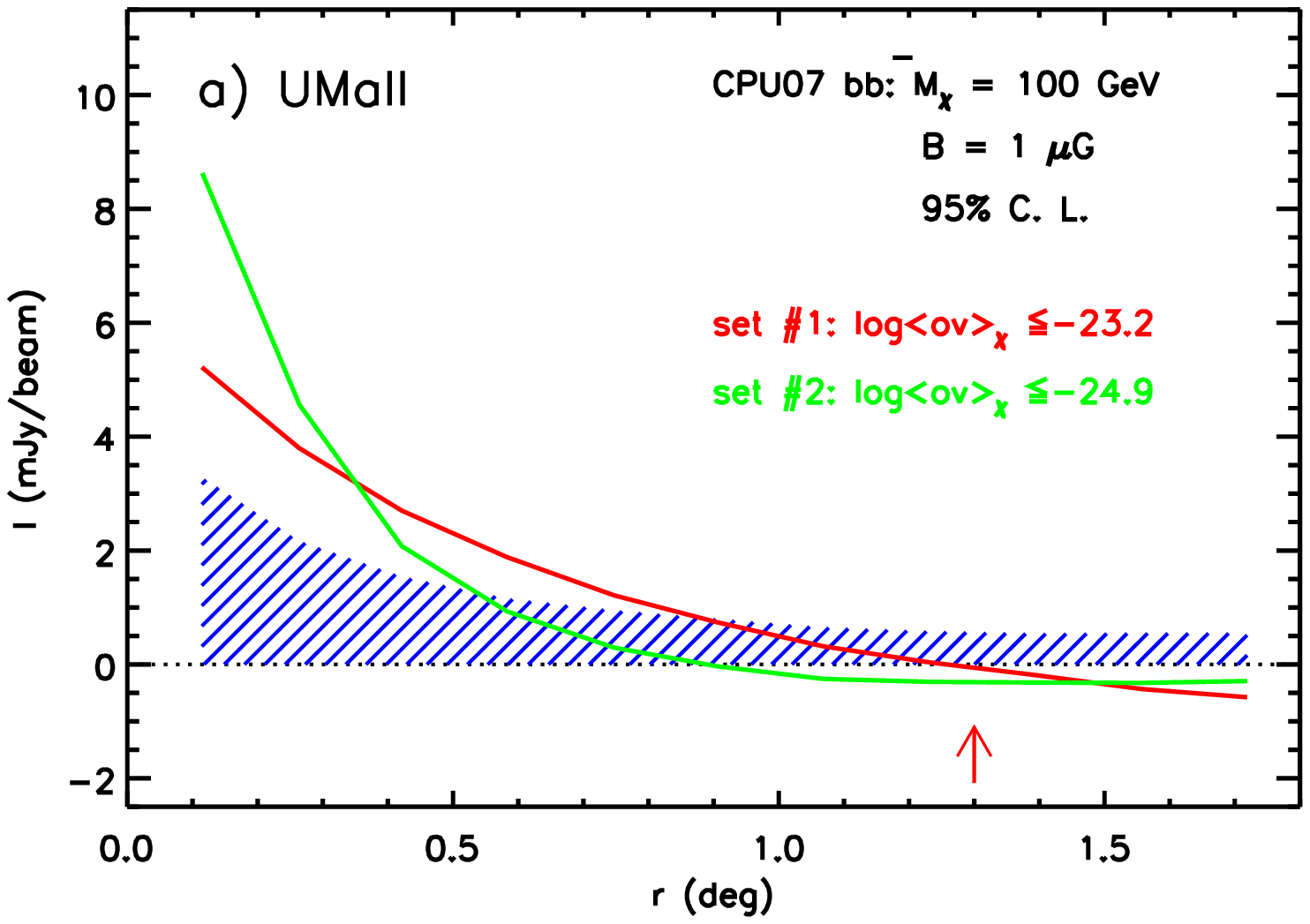}
\plotone{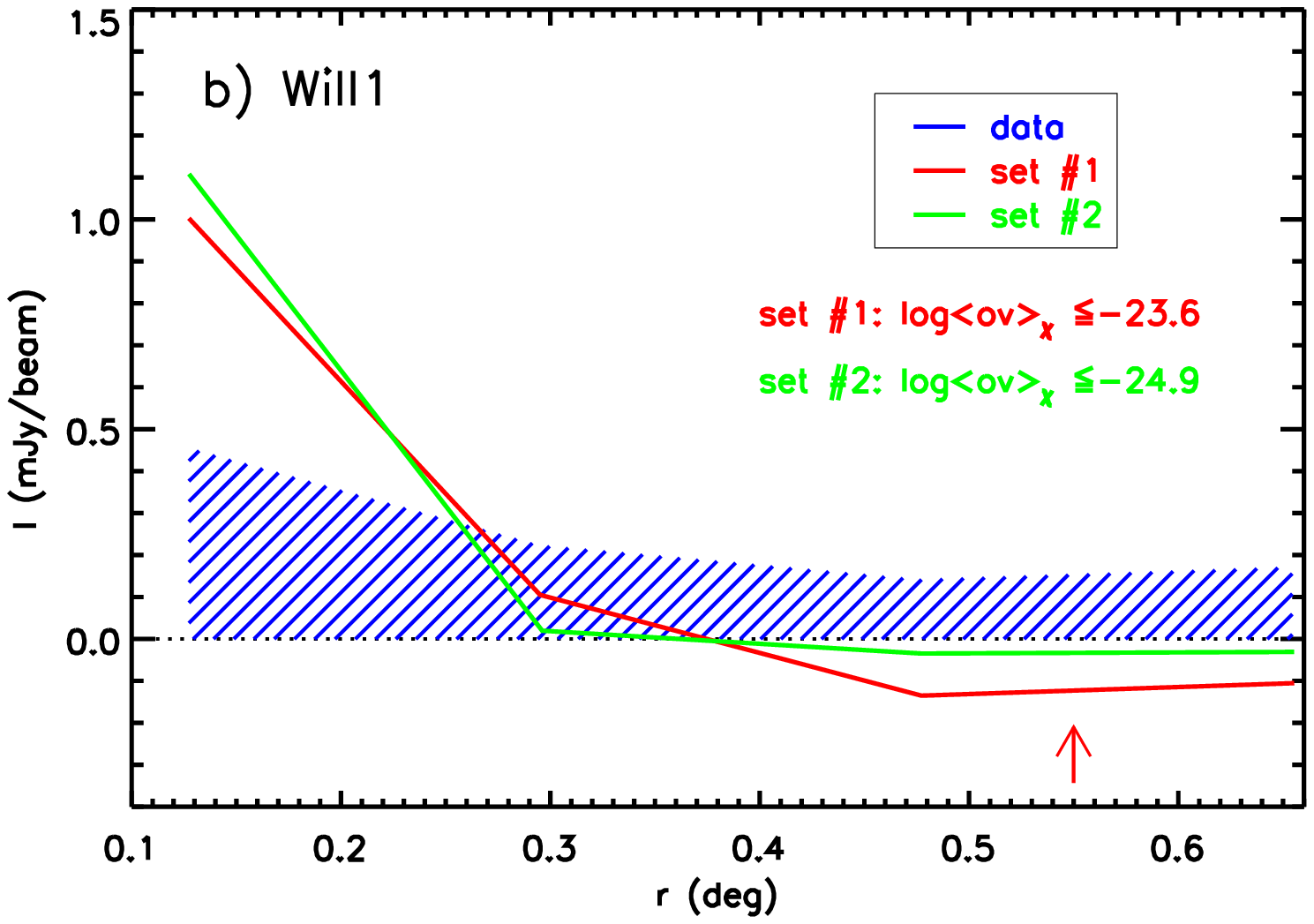}
\caption{Upper limits on \sigv\ from a) UMaII and b) Will1 for the CPU07 models shown in Fig.~\ref{fig:pred}, scaled radially using $r_h$ for that dSph (Table~\ref{tab:basic}). In both panels, the hatched (blue in the online version) region shows the radial profile uncertainties from Fig.~\ref{fig:prof}. The black (red in the online version) and grey (green in the online version) lines show the CPU07 halo profile shapes for ``set \#1" and ``set \#2" respectively, processed through our pipeline to account for filtering by our baselining procedure and scaled in intensity to be consistent with the profile uncertainties at 95\% confidence. Only uncertainties on profile points to the left of the vertical arrow are included in the fit. $\log\sigv$ given on the right-hand side of the panels is the corresponding limit implied by the CPU07 models for a thermal relic, in $\mathrm{cm^3s^{-1}}$.  ``Set \#2" is likely to be most appropriate for dSphs (\S\ref{discuss:cpu07:limits}).\\
(A color version of this figure is available in the online version of the Journal.)}
\label{fig:otherlim}
%\end{figure}
\end{figure}

 \begin{figure}
\epsscale{1.1}
\plotone{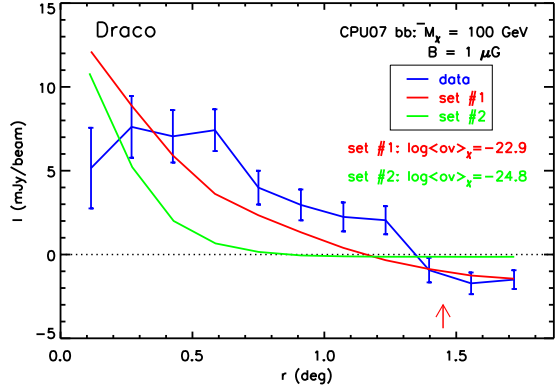}
\caption{Best fitting processed ``set \#1" and ``set \#2" CPU07 predictions to the radial profile computed for the Draco field. The solid line with error bars shows the radial surface brightness profile points as in Fig.~\ref{fig:prof}. The dotted black (red in the online version) and grey (green in the online version) lines show the best fitting``set \#1" and ``set \#2" halo profiles, processed through our pipeline to account for filtering by our baselining procedure. Only profile points to the left of the vertical arrow are included in the fit. The corresponding $\log\sigv$ is given on the right-hand side of the panels, in $\mathrm{cm^3s^{-1}}$. Note that the detected emission likely stems from foregrounds and not WIMP annihilations; see text for details. \\
(A color version of this figure is available in the online journal.)
\label{fig:dracolim}
}
\end{figure}

\subsection{Magnetic Fields in dSphs}
\label{discuss:bfields}

Unlike their counterparts at $\gamma$-ray energies, WIMP annihilations in dSphs are detectable in the radio only if these systems harbor large-scale magnetic fields. CPU07 note that the dependence of the signal on the magnetic field strength is not trivial because it affects both the particle propagation and the synchrotron emission, but the sensitivity of their fiducial models scales roughly as $\sigv \propto B$ (see their fig.~14). The magnetic field strength therefore has an important influence on the detectability of the predicted signal.  Here we discuss the plausibility that the dSphs considered here have turbulent magnetic fields with strengths $B\sim1\,\mu$G, as assumed in the fiducial CPU07 models of Fig.~\ref{fig:pred}. 
%However, because magnetic fields are detected through their effects on the ISM \citep[e.g.][]{beck11}, their strengths are unconstrained in largely ISM-free early-type galaxies such as the dSphs.

Direct measurements of $B$ are difficult to obtain \citep[see][for a review]{beck11}.  A standard technique is to search for radio emission attributable to synchrotron radiation, and to use the equipartition theorem to estimate the field strength.  Under modest assumptions regarding the synchrotron spectral index and the degree of polarization, one can infer the total magnetic field strength. Using this approach, turbulent magnetic field strengths of $\sim10\,\mu$G have been found in some actively star-forming, present-day dwarf irregular (dIrr) galaxies \citep{chyzy00,chyzy03,kepley10,kepley11}, while those in more quiescent systems like the LMC \citep{klein89} are in the $\sim3-5\,\mu$G range \citep[see also][]{chyzy11}. However, the lack of a detectable ISM precludes using this approach in present-day dSphs, and $B$ therein is largely unconstrained.

In principle, an estimate of the line-of-sight integral of the magnetic
fields and electron densities in dSphs can be obtained by measuring the
Faraday rotation of the polarization angle of the polarized emission
from background galaxies along the line-of-sight, i.e., the rotation
measure. This has been done for the LMC \citep{gaensler05}, where
they used polarized background galaxies with an areal density of 2.2 per
square degree, measured with a sensitivity of $0.2\,$mJy/beam at $1.4\,$GHz.
This magnetic field constraint underpinned the DM annihilation analysis
of \citet{siffert11}. Such a measurement could be done for dSphs, though it remains challenging. The most complete all-sky RM compilation \citep{oppermann12} has about 1 source per square degree and reconstructs the RM field with a resolution of $\sim30\arcmin$, with a high degree of non-uniformity, making it marginal for our purposes.

%In principle, an estimate of the line-of-sight integral of the magnetic fields and electron densities in dSphs can be obtained by measuring the Faraday rotation of the polarization angle of the polarized emission from background galaxies along the line-of-sight, i.e., the rotation measure.  This measurement is challenging since most radio galaxies have polarized fractions of $\sim2\%$, and we are looking for a change in rotation measure of order a few. This requires well-calibrated measurements at widely separated frequencies to determine the phase rotation.  Very low frequency telescopes can measure multiple wrappings within their bandwidth, but lack the requisite sensitivity at present (e.g., LOFAR, PAPER, MWA). It is therefore unlikely that Faraday rotation studies will constrain dSph magnetic fields in the near future.

Notwithstanding the difficulties in measuring $B$ in dSphs today, evidence for similar evolutionary histories between dIrrs and dSphs imply that the latter had magnetic fields in the past. The broadband photometric properties and resolved stellar populations of nearby dwarfs suggest that dIrrs and gas-poor dSphs had similar star formation histories until a few Gyr ago \citep{calura08,weisz11}. Numerical simulations indicate that ``tidal stirring", a combination of tidal and ram-pressure stripping by the Milky Way halo, is effective at transforming dIrrs into (classical) dSphs (see \citealt{mayer10} for a review), again suggesting a common origin for these two classes.   If dSphs and dIrrs share a common origin, then dSphs may once have had turbulent magnetic fields of order $5-10\,\mu$G as well. How this early magnetic field evolved after star formation ceased and their gas was stripped remains an open question. 

It is nonetheless plausible that a tenuous, ionised ISM that has insofar escaped detection the dSphs sustains a weak magnetic field. For example, Draco could contain interstellar gas with a mass that is $\sim10\%$ of that in its stars and still satisfy available upper bounds on its diffuse H$\alpha$ intensity \citep{gallagher03}. On the other hand, if we invoke equipartitition in UMaII and Will1 and make standard assumptions regarding the synchrotron path length ($l=r_h$ from Table~\ref{tab:basic}) and the proton-to-electron ratio ($K_o=100$), the upper limits on the peak ``set \#1" and ``set \#2" flux densities in Fig.~\ref{fig:otherlim} would allow $B \sim 1\,\mu$G in these systems \citep{beck05}. If the radio flux from low-luminosity galaxies is suppressed relative to that in $L_*$ galaxies (e.g.\ Bell 2003), $B$ could be larger.
We conclude that although detailed modeling of the magnetic field evolution in dSphs is required to place quantitative limits on their most likely present-day strengths, it is plausible that these systems harbor a turbulent magnetic field with $B=1\,\mu$G as invoked by CPU07. 

%Turning the problem around, does invoking B=1muG in Draco violate our upper limits on its radio flux? If there is equipartition between the energy densities in cosmic rays and in the magnetic field \citep{beck05}, our upper limit on the radio flux at the centre of Draco's subtracted profile translates into an upper limit\footnote{We assume a synchrotron pathlength equal to Draco's Holmberg radius (Table~\ref{tab:basic}), a proton-to-electron ratio $K_o=100$, a synchrotron spectral index $\alpha = - 0.7$ and a negligible thermal fraction. } of $B \lesssim 1\,\mu$G.  If the radio flux from low-luminosity galaxies is suppressed relative to that in $\simL_*$ galaxies (e.g.\ Bell 2003), our upper limits are less stringent.  Finally, we note that if Draco falls on the radio-FIR correlation as suggested by \citet{chyzy11} for dIrrs, the upper limit on the IRAS $60\,\mu$m flux of $F_{60} < 0.072\,$Jy \citep{lisenfeld07} implies a $1.4\,$GHz upper limit that is a few times higher than our average upper limit across Draco's optical disk. 

\subsection{The Way Forward}
\label{discuss:future}

Comparisons between the upper limits implied by the UMaII and Will1 surface brightness profiles and the preferred ``set \#2" CPU07 models in Fig.~\ref{fig:pred} require that $\log\sigv\ \lesssim -25$  for each system, which rivals the constraints obtained by Fermi-LAT for individual dSphs (\S\ref{discuss:cpu07}). It is therefore clear that deep radio observations of dSphs are highly complementary to WIMP searches at higher energies. We now discuss potential avenues for improving the constraints on \sigv\ from radio searches of dSphs relative to the results presented here:  mapping larger, offset fields (\S\ref{discuss:fmap}), carrying out a joint analysis of a large sample (\S\ref{discuss:fsample}), and exploring a variety of WIMP annihilation models (\S\ref{discuss:fmodel}).

\subsubsection{Mapping Larger, Offset dSph Fields}
\label{discuss:fmap}
While the constraints on \sigv\ from CPU07's ``set \#2" models and our data are competitive with those from the two-year Fermi-LAT data, the limits from ``set \#1" are considerably weaker (\S\ref{discuss:cpu07}). As explained in \S\ref{halo:bias}, this is due in part to the filtering effect of our baselining procedure on predicted halos with sizes comparable to the map size. This effect can be mitigated by mapping larger regions around each dSph: comparing the input and processed halo profiles derived in \S\ref{halo:bias}, we estimate that maps extending to $\sim20$ half-light radii $r_h$ (Table~\ref{tab:basic}) would be sufficient to mitigate the filtering of flux from CPU07's ``set \#1" prediction, and thus throughout the diffusion model parameter space relevant to the dSphs (\S\ref{discuss:cpu07:limits}). In the extreme case of ``set \#1" and the UMaII field, the larger maps proposed here would increase our sensitivity of \sigv\ by $\sim50\%$ (c.f. Fig.~\ref{fig:bias}). While costly in observing time (c.f.\ Table~\ref{tab:obs}), obtaining larger maps for several dSphs is nonetheless feasible with the GBT. For example, we estimate that a $6^\circ \times 6^\circ$ map of the same quality as those presented here -- which would probe the radio halo of Draco out to $\sim20r_h$ -- could be obtained with the GBT in $\sim$30 hours. 

As also discussed in \S\ref{halo:bias}, it is advantageous to offset the fields relative to the dSph centroids in order to minimize the curvature about the map center introduced by our baselining procedure. In the particular case of Draco, we estimate that an offset of $\sim5r_h$ is sufficient, with the added bonus of probing even farther into the putative halo of this dSph on one side. We note that larger, offset maps of Draco and Coma represent the most promising approach for mitigating the limitations imposed by foregrounds in the present study (c.f.\ \S\ref{discuss:foregrounds}), since in the absence of curvature bias (c.f.\ \S\ref{halo:bias}), foregrounds and annihilation halos should have different spatial distributions.  
%some of the limitations imposed by foregrounds in those fields, because the baselining procedure would be both less sensitive to foreground features in the map and less apt to correlate those features with the dSph centroids.

\subsubsection{A Larger Sample of dSphs}
\label{discuss:fsample}
In analogy to joint dSph analyses at $\gamma$-ray energies \citep{ackermann11,geringer-sameth11}, we anticipate that combining subtracted Stokes I maps for several dSph fields  will improve our constraints on \sigv. The sensitivity of our subtracted maps is limited by both systematic mapping errors and Galactic foregrounds ($\sigma_{map}$ and $\sigma_{ast}$ in Table~\ref{tab:noise}, respectively): particularly because of the latter issue (c.f.\ \S\ref{discuss:foregrounds}), it is unlikely that deeper observations of a single target will improve upon the \sigv\ limits presented here. However, our constraint on \sigv\ should tighten when a larger sample of dSphs is jointly considered, enabling detailed, simultaneous modelling of extended annihilation halos and the Galactic foreground (c.f. \S\ref{discuss:foregrounds:bayesian}). There is certainly potential for enlarging the dSph sample presented here with additional GBT observations: all but one of the 10 dSphs considered by \citet{ackermann11} are accessible to the GBT, as are several other Galactic dSphs from the recent compilation of \citet{mcconnachie12}.  

A compromise between sample size and telescope time investment would be to target the subset of accessible dSphs with the highest predicted annihilation fluxes. In the particular case of the GBT, we estimate that a 150-hour survey program could obtain maps extending to $\sim20r_h$ of the dSphs with the brightest putative radio halos, with J-values in the range $18.4 \lesssim \log(\mathrm{J,\,GeV\,cm^{-5}}) \lesssim 19.6$ (\citealt{ackermann11}; see also \citealt{charbonnier11}): Segue 1, UMaII, Coma, Draco, Ursa Minor, and Sculptor. We expect to be able to map all of these high-latitude fields to the same sensitivity as achieved in this paper. This sample is a good fit for the GBT, since other potential targets are expected to be an order of magnitude fainter than this lower bound \citep{ackermann11}. It is therefore feasible to nearly double the size of the sample considered here, while also obtaining sufficiently large maps to mitigate the filtering and curvature bias effects of our baselining procedure. Optimistically, one would hope to constrain \sigv\ in these 6 systems, thus tripling the number of targets analysed in this paper and enabling a joint foreground-annihilation halo analysis.

\subsubsection{Detailed dSph and WIMP Models}
\label{discuss:fmodel}
As emphasized in \S\ref{discuss:cpu07:limits}, our comparisons to the CPU07 models in Fig.~\ref{fig:pred} -- particularly their ``set \#2" -- are sufficient to estimate the magnitude of \sigv\ probed by our maps. Recall, however, that the models considered here adopt a single WIMP mass ($\mc = 100\,$GeV) and magnetic field strength ($B=1\,\mu$G), and only two sets of diffusion parameters (``set \#1" and ``set \#2"). Our deep radio observations therefore afford a more sophisticated analysis than the preliminary work presented here. In particular, our upper limits could be refined through comparisons to a suite of predictions that span the expected parameter space for dSph diffusion models, and encorporate J-values and their uncertainties appropriate for each dSph.  A preliminary application of such predictions to our UMaII data indicates that the resulting constraints on \sigv\ are comparable to the ``set \#2" results presented here (A. Natarajan et al., in preparation). 
 Together with an exploration of different decay channels and \mc, a detailed set of models would afford a thorough investigation of the range of \sigv\ that are allowed by our data given different assumptions about both dSph and WIMP properties.

\section{Conclusions}

Models by CPU07 predict that for plausible values of the turbulent magnetic field strength and charged particle propagation parameters in the Draco dSph, WIMP annihilations in its dark matter halo will produce a degree-scale synchrotron radio halo that is accessible to current single-dish facilities. We present deep GBT observations at $1.4\,$ GHz of a total of $40.5\arcdeg^2$ around the Draco, UMaII, Coma, and Will1 dSphs to detect this annihilation signature. 

We search for extended radio emission associated with the dSphs at sensitivities $\sigma_{sub} \lesssim 7\,$mJy/beam (\S\ref{data:gbt}) by subtracting discrete sources in the Stokes I maps using the NVSS catalog. The subtracted map noise is not limited by our discrete source subtraction; it is therefore possible to map well below the nominal confusion limit with the GBT when the NVSS is used to remove background sources. We obtained near-concurrent observations of the Draco field with the VLA (\S\ref{data:vla}), and use them to demonstrate that variability in the discrete source population has a negligible impact on our results (\S\ref{halo:var}).

For each subtracted Stokes I map, we compute radial profiles about the dSph at the map center, jacknifing the data in observation time, telescope elevation and telescope PA to determine robust uncertainties on each profile point (\S\ref{halo:prof}). We find that $I(r) \neq 0$ with high significance in the Draco and Coma fields. The depression in the Coma field almost certainly stems from foregrounds. While the emission profile computed from the Draco field resembles that expected for an extended halo, several lines of evidence suggest that it too stems from foregrounds (\S\ref{discuss:cpu07:draco}). 
Indeed, our subtracted maps probe foregrounds at unprecedented sensitivity and resolution (\S\ref{discuss:foregrounds}): we find that the standard deviation in our maps attributable to foregrounds is $1.8\,\mathrm{mJy/beam} \leq \sigma_{ast} \leq 5.7\,\mathrm{mJy/beam}$, where $\sigma_{ast}$ roughly anti-correlates with the galactic latitude range  $35\arcdeg \leq b \leq 84\arcdeg$ of the maps. Because of their strong foreground contamination, we do not attempt to constrain \sigv\ from the Draco or Coma fields.

We find no statistically significant features in the UMaII and Will1 profiles, suggesting that these fields are devoid of both foreground contamination and WIMP annihilations. Assuming that the emission from a putative radio halo in these dSphs is consistent with zero, we use the uncertainties on the profile points in those fields to place limits on the annihilation cross-section \sigv\ in the context of the CPU07 models for Draco (\S\ref{discuss:cpu07:limits}), which we scale radially according to the stellar distribution sizes of the other dSphs.  We compare the fiducial CPU07 models with $M_\chi = 100\,$GeV annihilating into $b\bar{b}$ final states, $B= 1\,\mu$G and two sets of charged particle propagation parameters, and account for the filtering of those models by our baselining procedure (\S\ref{halo:bias}). We argue that CPU07's ``set \#2" model is most appropriate for dSphs, and that the resulting annihilation flux predictions are likely valid to within an order of magnitude.

In the context of the CPU07 models, our upper limits from the UMaII and Wil1 fields imply $\log(\sigv, \mathrm{\,cm^3\,s^{-1}}) \lesssim -25$ for the preferred ``set \#2" propagation parameters (\S\ref{discuss:cpu07:limits}). This constraint is comparable to the limits for individual dSphs obtained at $\gamma$-ray energies from the two-year Fermi-LAT data. We discuss three potential avenues for improving the limits on \sigv\ from deep radio observations: mapping larger fields that are offset from the dSph optical centroids (\S\ref{discuss:fmap}),  carrying out a joint analysis of a larger sample of dSphs (\S\ref{discuss:fsample}),  and applying a more sophisticated suite of models to the data (\S\ref{discuss:fmodel}). 
%For at least some of the models considered here, we may gain up to an order of magnitude in sensitivity by pursuing these avenues. 
We therefore conclude that deep radio observations are highly complementary to indirect WIMP searches at higher energies, and have the potential to probe \sigvth\ expected for a thermal relic.

\acknowledgements

\nraoack\
This project was initiated while K.~S.\ and J.~A.\ were Jansky Fellows of the National Radio Astronomy Observatory. K.~S.\ acknowledges support from the National Science and Engineering Research Council of Canada. We thank S.\ Srikanth for providing the GBT $\nu=1.4\,$GHz beam model used in this analysis, A.\ Natarajan for helpful discussions regarding WIMP annihilation models and the anonymous referee whose comments helped clarify the interpretation of our data.

{\it Facilities:} \facility{NRAO: GBT, VLA}

\bibliography{refs}

\begin{thebibliography}{87}
\expandafter\ifx\csname natexlab\endcsname\relax\def\natexlab#1{#1}\fi

\bibitem[{{Abazajian} \& {Harding}(2012)}]{abazajian12a}
{Abazajian}, K.~N. \& {Harding}, J.~P. 2012, JCAP, 1, 41

\bibitem[{{Abazajian} \& {Kaplinghat}(2012)}]{abazajian12b}
{Abazajian}, K.~N. \& {Kaplinghat}, M. 2012, \prd, 86, 083511

\bibitem[{{Abdo} {et~al.}(2010{\natexlab{a}}){Abdo}, {Ackermann}, {Ajello},
  {Atwood}, {Baldini}, {Ballet}, {Barbiellini}, {Bastieri}, {Bechtol},
  {Bellazzini}, {Berenji}, {Bloom}, {Bonamente}, {Borgland}, {Bouvier},
  {Bregeon}, {Brez}, {Brigida}, {Bruel}, {Burnett}, {Buson}, {Caliandro},
  {Cameron}, {Caraveo}, {Carrigan}, {Casandjian}, {Cecchi}, {{\c C}elik},
  {Chekhtman}, {Chiang}, {Ciprini}, {Claus}, {Cohen-Tanugi}, {Conrad},
  {Dermer}, {de Angelis}, {de Palma}, {Digel}, {Do Couto E Silva}, {Drell},
  {Drlica-Wagner}, {Dubois}, {Dumora}, {Edmonds}, {Essig}, {Farnier},
  {Favuzzi}, {Fegan}, {Focke}, {Fortin}, {Frailis}, {Fukazawa}, {Funk},
  {Fusco}, {Gargano}, {Gasparrini}, {Gehrels}, {Germani}, {Giglietto},
  {Giordano}, {Glanzman}, {Godfrey}, {Grenier}, {Grove}, {Guillemot},
  {Guiriec}, {Gustafsson}, {Hadasch}, {Harding}, {Horan}, {Hughes}, {Jackson},
  {J{\'o}hannesson}, {Johnson}, {Johnson}, {Johnson}, {Kamae}, {Katagiri},
  {Kataoka}, {Kawai}, {Kerr}, {Kn{\"o}dlseder}, {Kuss}, {Lande}, {Latronico},
  {Llena Garde}, {Longo}, {Loparco}, {Lott}, {Lovellette}, {Lubrano}, {Makeev},
  {Mazziotta}, {McEnery}, {Meurer}, {Michelson}, {Mitthumsiri}, {Mizuno},
  {Moiseev}, {Monte}, {Monzani}, {Morselli}, {Moskalenko}, {Murgia}, {Nolan},
  {Norris}, {Nuss}, {Ohsugi}, {Omodei}, {Orlando}, {Ormes}, {Ozaki}, {Paneque},
  {Panetta}, {Parent}, {Pelassa}, {Pepe}, {Pesce-Rollins}, {Piron},
  {Rain{\`o}}, {Rando}, {Razzano}, {Reimer}, {Reimer}, {Reposeur}, {Ripken},
  {Ritz}, {Rodriguez}, {Roth}, {Sadrozinski}, {Sander}, {Parkinson}, {Scargle},
  {Schalk}, {Sellerholm}, {Sgr{\`o}}, {Siskind}, {Smith}, {Smith}, {Spandre},
  {Spinelli}, {Starck}, {Strickman}, {Suson}, {Tajima}, {Takahashi}, {Tanaka},
  {Thayer}, {Thayer}, {Tibaldo}, {Torres}, {Uchiyama}, {Usher}, {Vasileiou},
  {Vilchez}, {Vitale}, {Waite}, {Wang}, {Winer}, {Wood}, {Ylinen}, {Ziegler},
  \& {Fermi LAT Collaboration}}]{abdo10a}
{Abdo}, A.~A. {et~al.} 2010{\natexlab{a}}, Physical Review Letters, 104, 091302

\bibitem[{{Abdo} {et~al.}(2010{\natexlab{b}}){Abdo}, {Ackermann}, {Ajello},
  {Atwood}, {Baldini}, {Ballet}, {Barbiellini}, {Bastieri}, {Bechtol},
  {Bellazzini}, {Berenji}, {Bloom}, {Bonamente}, {Borgland}, {Bregeon}, {Brez},
  {Brigida}, {Bruel}, {Burnett}, {Buson}, {Caliandro}, {Cameron}, {Caraveo},
  {Casandjian}, {Cecchi}, {Chekhtman}, {Cheung}, {Chiang}, {Ciprini}, {Claus},
  {Cohen-Tanugi}, {Conrad}, {de Angelis}, {de Palma}, {Digel}, {Silva},
  {Drell}, {Drlica-Wagner}, {Dubois}, {Dumora}, {Farnier}, {Favuzzi}, {Fegan},
  {Focke}, {Fortin}, {Frailis}, {Fukazawa}, {Fusco}, {Gargano}, {Gehrels},
  {Germani}, {Giebels}, {Giglietto}, {Giordano}, {Glanzman}, {Godfrey},
  {Grenier}, {Grove}, {Guillemot}, {Guiriec}, {Gustafsson}, {Harding}, {Hays},
  {Horan}, {Hughes}, {Jackson}, {Jeltema}, {J{\'o}hannesson}, {Johnson},
  {Johnson}, {Johnson}, {Kamae}, {Katagiri}, {Kataoka}, {Kerr},
  {Kn{\"o}dlseder}, {Kuss}, {Lande}, {Latronico}, {Lemoine-Goumard}, {Longo},
  {Loparco}, {Lott}, {Lovellette}, {Lubrano}, {Madejski}, {Makeev},
  {Mazziotta}, {McEnery}, {Meurer}, {Michelson}, {Mitthumsiri}, {Mizuno},
  {Moiseev}, {Monte}, {Monzani}, {Moretti}, {Morselli}, {Moskalenko}, {Murgia},
  {Nolan}, {Norris}, {Nuss}, {Ohsugi}, {Omodei}, {Orlando}, {Ormes}, {Paneque},
  {Panetta}, {Parent}, {Pelassa}, {Pepe}, {Pesce-Rollins}, {Piron}, {Porter},
  {Profumo}, {Rain{\`o}}, {Rando}, {Razzano}, {Reimer}, {Reimer}, {Reposeur},
  {Ritz}, {Rodriguez}, {Roth}, {Sadrozinski}, {Sander}, {Saz Parkinson},
  {Scargle}, {Schalk}, {Sellerholm}, {Sgr{\`o}}, {Siskind}, {Smith}, {Smith},
  {Spandre}, {Spinelli}, {Strickman}, {Suson}, {Takahashi}, {Takahashi},
  {Tanaka}, {Thayer}, {Thayer}, {Thompson}, {Tibaldo}, {Torres}, {Tramacere},
  {Uchiyama}, {Usher}, {Vasileiou}, {Vilchez}, {Vitale}, {Waite}, {Wang},
  {Winer}, {Wood}, {Ylinen}, {Ziegler}, {Bullock}, {Kaplinghat}, {Martinez}, \&
  {Fermi-LAT Collaboration}}]{abdo10}
---. 2010{\natexlab{b}}, \apj, 712, 147

\bibitem[{{Abdo} {et~al.}(2010{\natexlab{c}}){Abdo}, {Ackermann}, {Ajello},
  {Baldini}, {Ballet}, {Barbiellini}, {Bastieri}, {Bechtol}, {Bellazzini},
  {Berenji}, {Blandford}, {Bloom}, {Bonamente}, {Borgland}, {Bouvier},
  {Bregeon}, {Brez}, {Brigida}, {Bruel}, {Burnett}, {Buson}, {Caliandro},
  {Cameron}, {Caraveo}, {Carrigan}, {Casandjian}, {Cecchi}, {{\c C}elik},
  {Chekhtman}, {Cheung}, {Chiang}, {Ciprini}, {Claus}, {Cohen-Tanugi},
  {Conrad}, {Cutini}, {Dermer}, {de Angelis}, {de Palma}, {Digel}, {Silva},
  {Drell}, {Dubois}, {Dumora}, {Edmonds}, {Farnier}, {Favuzzi}, {Fegan},
  {Focke}, {Fortin}, {Frailis}, {Fukazawa}, {Fusco}, {Gargano}, {Gasparrini},
  {Gehrels}, {Germani}, {Giglietto}, {Giordano}, {Glanzman}, {Godfrey},
  {Grove}, {Guillemot}, {Guiriec}, {Gustafsson}, {Hadasch}, {Harding}, {Horan},
  {Hughes}, {Johnson}, {Johnson}, {Kamae}, {Katagiri}, {Kataoka}, {Kawai},
  {Kerr}, {Kn{\"o}dlseder}, {Kuss}, {Lande}, {Latronico}, {Llena Garde},
  {Longo}, {Loparco}, {Lott}, {Lovellette}, {Lubrano}, {Makeev}, {Mazziotta},
  {McEnery}, {Meurer}, {Michelson}, {Mitthumsiri}, {Mizuno}, {Monte},
  {Monzani}, {Morselli}, {Moskalenko}, {Murgia}, {Nolan}, {Norris}, {Nuss},
  {Ohsugi}, {Omodei}, {Orlando}, {Ormes}, {Paneque}, {Panetta}, {Parent},
  {Pelassa}, {Pepe}, {Pesce-Rollins}, {Piron}, {Rain{\`o}}, {Rando}, {Reimer},
  {Reimer}, {Reposeur}, {Rodriguez}, {Roth}, {Sadrozinski}, {Sander}, {Saz
  Parkinson}, {Scargle}, {Sellerholm}, {Sgr{\`o}}, {Siskind}, {Smith},
  {Spandre}, {Spinelli}, {Starck}, {Strickman}, {Suson}, {Takahashi}, {Tanaka},
  {Thayer}, {Thayer}, {Torres}, {Uchiyama}, {Usher}, {Vasileiou}, {Vilchez},
  {Vitale}, {Waite}, {Wang}, {Winer}, {Wood}, {Ylinen}, {Zaharijas}, \&
  {Ziegler}}]{abdo10b}
---. 2010{\natexlab{c}}, JCAP, 4, 14

\bibitem[{{Abramowski} {et~al.}(2012){Abramowski}, {Acero}, {Aharonian},
  {Akhperjanian}, {Anton}, {Balzer}, {Barnacka}, {Barres de Almeida},
  {Becherini}, {Becker}, {Behera}, {Bernl{\"o}hr}, {Birsin}, {Biteau},
  {Bochow}, {Boisson}, {Bolmont}, {Bordas}, {Brucker}, {Brun}, {Brun}, {Bulik},
  {B{\"u}sching}, {Carrigan}, {Casanova}, {Cerruti}, {Chadwick}, {Charbonnier},
  {Chaves}, {Cheesebrough}, {Clapson}, {Coignet}, {Cologna}, {Conrad},
  {Dalton}, {Daniel}, {Davids}, {Degrange}, {Deil}, {Dickinson},
  {Djannati-Ata{\"i}}, {Domainko}, {Drury}, {Dubus}, {Dutson}, {Dyks}, {Dyrda},
  {Egberts}, {Eger}, {Espigat}, {Fallon}, {Farnier}, {Fegan}, {Feinstein},
  {Fernandes}, {Fiasson}, {Fontaine}, {F{\"o}rster}, {F{\"u}{\ss}ling},
  {Gallant}, {Gast}, {G{\'e}rard}, {Gerbig}, {Giebels}, {Glicenstein},
  {Gl{\"u}ck}, {Goret}, {G{\"o}ring}, {H{\"a}ffner}, {Hague}, {Hampf},
  {Hauser}, {Heinz}, {Heinzelmann}, {Henri}, {Hermann}, {Hinton}, {Hoffmann},
  {Hofmann}, {Hofverberg}, {Holler}, {Horns}, {Jacholkowska}, {de Jager},
  {Jahn}, {Jamrozy}, {Jung}, {Kastendieck}, {Katarzy{\'n}ski}, {Katz},
  {Kaufmann}, {Keogh}, {Khangulyan}, {Kh{\'e}lifi}, {Klochkov}, {Klu{\'z}niak},
  {Kneiske}, {Komin}, {Kosack}, {Kossakowski}, {Laffon}, {Lamanna}, {Lennarz},
  {Lohse}, {Lopatin}, {Lu}, {Marandon}, {Marcowith}, {Masbou}, {Maurin},
  {Maxted}, {Mayer}, {McComb}, {Medina}, {M{\'e}hault}, {Moderski}, {Moulin},
  {Naumann}, {Naumann-Godo}, {de Naurois}, {Nedbal}, {Nekrassov}, {Nguyen},
  {Nicholas}, {Niemiec}, {Nolan}, {Ohm}, {de O{\~n}a Wilhelmi}, {Opitz},
  {Ostrowski}, {Oya}, {Panter}, {Paz Arribas}, {Pedaletti}, {Pelletier},
  {Petrucci}, {Pita}, {P{\"u}hlhofer}, {Punch}, {Quirrenbach}, {Raue},
  {Rayner}, {Reimer}, {Reimer}, {Renaud}, {de los Reyes}, {Rieger}, {Ripken},
  {Rob}, {Rosier-Lees}, {Rowell}, {Rudak}, {Rulten}, {Ruppel}, {Sahakian},
  {Sanchez}, {Santangelo}, {Schlickeiser}, {Sch{\"o}ck}, {Schulz}, {Schwanke},
  {Schwarzburg}, {Schwemmer}, {Sheidaei}, {Skilton}, {Sol}, {Spengler},
  {Stawarz}, {Steenkamp}, {Stegmann}, {Stinzing}, {Stycz}, {Sushch}, {Szostek},
  {Tavernet}, {Terrier}, {Tluczykont}, {Valerius}, {van Eldik}, {Vasileiadis},
  {Venter}, {Vialle}, {Viana}, {Vincent}, {V{\"o}lk}, {Volpe}, {Vorobiov},
  {Vorster}, {Wagner}, {Ward}, {White}, {Wierzcholska}, {Zacharias}, {Zajczyk},
  {Zdziarski}, {Zech}, {Zechlin}, \&
  {H.~E.~S.~S.~Collaboration}}]{abramowski12}
{Abramowski}, A. {et~al.} 2012, \apj, 750, 123

\bibitem[{{Ackermann} {et~al.}(2011){Ackermann}, {Ajello}, {Albert}, {Atwood},
  {Baldini}, {Ballet}, {Barbiellini}, {Bastieri}, {Bechtol}, {Bellazzini},
  {Berenji}, {Blandford}, {Bloom}, {Bonamente}, {Borgland}, {Bregeon},
  {Brigida}, {Bruel}, {Buehler}, {Burnett}, {Buson}, {Caliandro}, {Cameron},
  {Ca{\~n}adas}, {Caraveo}, {Casandjian}, {Cecchi}, {Charles}, {Chekhtman},
  {Chiang}, {Ciprini}, {Claus}, {Cohen-Tanugi}, {Conrad}, {Cutini}, {de
  Angelis}, {de Palma}, {Dermer}, {Digel}, {Do Couto E Silva}, {Drell},
  {Drlica-Wagner}, {Falletti}, {Favuzzi}, {Fegan}, {Ferrara}, {Fukazawa},
  {Funk}, {Fusco}, {Gargano}, {Gasparrini}, {Gehrels}, {Germani}, {Giglietto},
  {Giordano}, {Giroletti}, {Glanzman}, {Godfrey}, {Grenier}, {Guiriec},
  {Gustafsson}, {Hadasch}, {Hayashida}, {Hays}, {Hughes}, {Jeltema},
  {J{\'o}hannesson}, {Johnson}, {Johnson}, {Kamae}, {Katagiri}, {Kataoka},
  {Kn{\"o}dlseder}, {Kuss}, {Lande}, {Latronico}, {Lionetto}, {Llena Garde},
  {Longo}, {Loparco}, {Lott}, {Lovellette}, {Lubrano}, {Madejski}, {Mazziotta},
  {McEnery}, {Mehault}, {Michelson}, {Mitthumsiri}, {Mizuno}, {Monte},
  {Monzani}, {Morselli}, {Moskalenko}, {Murgia}, {Naumann-Godo}, {Norris},
  {Nuss}, {Ohsugi}, {Okumura}, {Omodei}, {Orlando}, {Ormes}, {Ozaki},
  {Paneque}, {Parent}, {Pesce-Rollins}, {Pierbattista}, {Piron}, {Pivato},
  {Porter}, {Profumo}, {Rain{\`o}}, {Razzano}, {Reimer}, {Reimer}, {Ritz},
  {Roth}, {Sadrozinski}, {Sbarra}, {Scargle}, {Schalk}, {Sgr{\`o}}, {Siskind},
  {Spandre}, {Spinelli}, {Strigari}, {Suson}, {Tajima}, {Takahashi}, {Tanaka},
  {Thayer}, {Thayer}, {Thompson}, {Tibaldo}, {Tinivella}, {Torres}, {Troja},
  {Uchiyama}, {Vandenbroucke}, {Vasileiou}, {Vianello}, {Vitale}, {Waite},
  {Wang}, {Winer}, {Wood}, {Wood}, {Yang}, {Zimmer}, {Kaplinghat}, \&
  {Martinez}}]{ackermann11}
{Ackermann}, M. {et~al.} 2011, Physical Review Letters, 107, 241302

\bibitem[{{Ackermann} {et~al.}(2010){Ackermann}, {Ajello}, {Allafort},
  {Baldini}, {Ballet}, {Barbiellini}, {Bastieri}, {Bechtol}, {Bellazzini},
  {Blandford}, {Bloom}, {Bonamente}, {Borgland}, {Bouvier}, {Brandt},
  {Bregeon}, {Brigida}, {Bruel}, {Buehler}, {Buson}, {Caliandro}, {Cameron},
  {Caraveo}, {Carrigan}, {Casandjian}, {Cecchi}, {Charles}, {Chekhtman},
  {Cheung}, {Chiang}, {Ciprini}, {Claus}, {Cohen-Tanugi}, {Cominsky}, {Conrad},
  {de Angelis}, {de Palma}, {Silva}, {Drell}, {Drlica-Wagner}, {Dubois},
  {Dumora}, {Edmonds}, {Farnier}, {Favuzzi}, {Fegan}, {Frailis}, {Fukazawa},
  {Fusco}, {Gargano}, {Gasparrini}, {Gehrels}, {Germani}, {Giglietto},
  {Giordano}, {Glanzman}, {Godfrey}, {Grenier}, {Guiriec}, {Gustafsson},
  {Harding}, {Hayashida}, {Horan}, {Hughes}, {Jeltema}, {J{\'o}hannesson},
  {Johnson}, {Johnson}, {Kamae}, {Katagiri}, {Kataoka}, {Kn{\"o}dlseder},
  {Kuss}, {Lande}, {Latronico}, {Lee}, {Llena Garde}, {Longo}, {Loparco},
  {Lovellette}, {Lubrano}, {Madejski}, {Makeev}, {Mazziotta}, {Michelson},
  {Mitthumsiri}, {Mizuno}, {Moiseev}, {Monte}, {Monzani}, {Morselli},
  {Moskalenko}, {Murgia}, {Nolan}, {Norris}, {Nuss}, {Ohno}, {Ohsugi},
  {Omodei}, {Orlando}, {Ormes}, {Panetta}, {Pepe}, {Pesce-Rollins}, {Piron},
  {Porter}, {Profumo}, {Rain{\`o}}, {Razzano}, {Reposeur}, {Ritz}, {Rodriguez},
  {Roth}, {Sadrozinski}, {Sander}, {Scargle}, {Sgr{\`o}}, {Siskind}, {Smith},
  {Spandre}, {Spinelli}, {Starck}, {Strickman}, {Suson}, {Takahashi}, {Tanaka},
  {Thayer}, {Thayer}, {Tibaldo}, {Torres}, {Tosti}, {Usher}, {Vasileiou},
  {Vitale}, {Waite}, {Wang}, {Winer}, {Wood}, {Yang}, {Ylinen}, {Ziegler}, \&
  {Fermi LAT Collaboration}}]{ackermann10}
---. 2010, JCAP, 5, 25

\bibitem[{{Ackermann} {et~al.}(2012{\natexlab{a}}){Ackermann}, {Ajello},
  {Allafort}, {Baldini}, {Ballet}, {Bastieri}, {Bechtol}, {Bellazzini},
  {Berenji}, {Bloom}, {Bonamente}, {Borgland}, {Bouvier}, {Bregeon}, {Brigida},
  {Bruel}, {Buehler}, {Buson}, {Caliandro}, {Cameron}, {Caraveo}, {Casandjian},
  {Cecchi}, {Charles}, {Chekhtman}, {Cheung}, {Chiang}, {Cillis}, {Ciprini},
  {Claus}, {Cohen-Tanugi}, {Conrad}, {Cutini}, {de Palma}, {Dermer}, {Digel},
  {Silva}, {Drell}, {Drlica-Wagner}, {Favuzzi}, {Fegan}, {Fortin}, {Fukazawa},
  {Funk}, {Fusco}, {Gargano}, {Gasparrini}, {Germani}, {Giglietto}, {Giordano},
  {Glanzman}, {Godfrey}, {Grenier}, {Guiriec}, {Gustafsson}, {Hadasch},
  {Hayashida}, {Hays}, {Hughes}, {J{\'o}hannesson}, {Johnson}, {Kamae},
  {Katagiri}, {Kataoka}, {Kn{\"o}dlseder}, {Kuss}, {Lande}, {Longo}, {Loparco},
  {Lott}, {Lovellette}, {Lubrano}, {Madejski}, {Martin}, {Mazziotta},
  {McEnery}, {Michelson}, {Mizuno}, {Monte}, {Monzani}, {Morselli},
  {Moskalenko}, {Murgia}, {Nishino}, {Norris}, {Nuss}, {Ohno}, {Ohsugi},
  {Okumura}, {Omodei}, {Orlando}, {Ozaki}, {Parent}, {Persic}, {Pesce-Rollins},
  {Petrosian}, {Pierbattista}, {Piron}, {Pivato}, {Porter}, {Rain{\`o}},
  {Rando}, {Razzano}, {Reimer}, {Reimer}, {Ritz}, {Roth}, {Sbarra}, {Sgr{\`o}},
  {Siskind}, {Spandre}, {Spinelli}, {Stawarz}, {Strong}, {Takahashi}, {Tanaka},
  {Thayer}, {Tibaldo}, {Tinivella}, {Torres}, {Tosti}, {Troja}, {Uchiyama},
  {Vandenbroucke}, {Vianello}, {Vitale}, {Waite}, {Wood}, \&
  {Yang}}]{ackermann12e}
---. 2012{\natexlab{a}}, ArXiv e-prints

\bibitem[{{Ackermann} {et~al.}(2012{\natexlab{b}}){Ackermann}, {Ajello},
  {Allafort}, {Baldini}, {Ballet}, {Bastieri}, {Bechtol}, {Bellazzini},
  {Berenji}, {Bloom}, {Bonamente}, {Borgland}, {Bouvier}, {Bregeon}, {Brigida},
  {Bruel}, {Buehler}, {Buson}, {Caliandro}, {Cameron}, {Caraveo}, {Casandjian},
  {Cecchi}, {Charles}, {Chekhtman}, {Cheung}, {Chiang}, {Cillis}, {Ciprini},
  {Claus}, {Cohen-Tanugi}, {Conrad}, {Cutini}, {de Palma}, {Dermer}, {Digel},
  {Silva}, {Drell}, {Drlica-Wagner}, {Favuzzi}, {Fegan}, {Fortin}, {Fukazawa},
  {Funk}, {Fusco}, {Gargano}, {Gasparrini}, {Germani}, {Giglietto}, {Giordano},
  {Glanzman}, {Godfrey}, {Grenier}, {Guiriec}, {Gustafsson}, {Hadasch},
  {Hayashida}, {Hays}, {Hughes}, {J{\'o}hannesson}, {Johnson}, {Kamae},
  {Katagiri}, {Kataoka}, {Kn{\"o}dlseder}, {Kuss}, {Lande}, {Longo}, {Loparco},
  {Lott}, {Lovellette}, {Lubrano}, {Madejski}, {Martin}, {Mazziotta},
  {McEnery}, {Michelson}, {Mizuno}, {Monte}, {Monzani}, {Morselli},
  {Moskalenko}, {Murgia}, {Nishino}, {Norris}, {Nuss}, {Ohno}, {Ohsugi},
  {Okumura}, {Omodei}, {Orlando}, {Ozaki}, {Parent}, {Persic}, {Pesce-Rollins},
  {Petrosian}, {Pierbattista}, {Piron}, {Pivato}, {Porter}, {Rain{\`o}},
  {Rando}, {Razzano}, {Reimer}, {Reimer}, {Ritz}, {Roth}, {Sbarra}, {Sgr{\`o}},
  {Siskind}, {Spandre}, {Spinelli}, {Stawarz}, {Strong}, {Takahashi}, {Tanaka},
  {Thayer}, {Tibaldo}, {Tinivella}, {Torres}, {Tosti}, {Troja}, {Uchiyama},
  {Vandenbroucke}, {Vianello}, {Vitale}, {Waite}, {Wood}, \&
  {Yang}}]{ackermann12b}
---. 2012{\natexlab{b}}, \apj, 755, 164

\bibitem[{{Aguirre} {et~al.}(2011){Aguirre}, {Ginsburg}, {Dunham}, {Drosback},
  {Bally}, {Battersby}, {Bradley}, {Cyganowski}, {Dowell}, {Evans}, {Glenn},
  {Harvey}, {Rosolowsky}, {Stringfellow}, {Walawender}, \&
  {Williams}}]{aguirre11}
{Aguirre}, J.~E. {et~al.} 2011, \apjs, 192, 4

\bibitem[{{Aleksi{\'c}} {et~al.}(2011){Aleksi{\'c}}, {Alvarez}, {Antonelli},
  {Antoranz}, {Asensio}, {Backes}, {Barrio}, {Bastieri}, {Becerra
  Gonz{\'a}lez}, {Bednarek}, {Berdyugin}, {Berger}, {Bernardini}, {Biland},
  {Blanch}, {Bock}, {Boller}, {Bonnoli}, {Borla Tridon}, {Braun}, {Bretz},
  {Ca{\~n}ellas}, {Carmona}, {Carosi}, {Colin}, {Colombo}, {Contreras},
  {Cortina}, {Cossio}, {Covino}, {Dazzi}, {De Angelis}, {De Cea del Pozo}, {De
  Lotto}, {Delgado Mendez}, {Diago Ortega}, {Doert}, {Dom{\'{\i}}nguez},
  {Dominis Prester}, {Dorner}, {Doro}, {Elsaesser}, {Ferenc}, {Fonseca},
  {Font}, {Fruck}, {Garc{\'{\i}}a L{\'o}pez}, {Garczarczyk}, {Garrido},
  {Giavitto}, {Godinovi{\'c}}, {Hadasch}, {H{\"a}fner}, {Herrero},
  {Hildebrand}, {H{\"o}hne-M{\"o}nch}, {Hose}, {Hrupec}, {Huber}, {Jogler},
  {Klepser}, {Kr{\"a}henb{\"u}hl}, {Krause}, {La Barbera}, {Lelas}, {Leonardo},
  {Lindfors}, {Lombardi}, {L{\'o}pez}, {Lorenz}, {Makariev}, {Maneva},
  {Mankuzhiyil}, {Mannheim}, {Maraschi}, {Mariotti}, {Mart{\'{\i}}nez},
  {Mazin}, {Meucci}, {Miranda}, {Mirzoyan}, {Miyamoto}, {Mold{\'o}n},
  {Moralejo}, {Munar-Androver}, {Nieto}, {Nilsson}, {Orito}, {Oya}, {Paiano},
  {Paneque}, {Paoletti}, {Pardo}, {Paredes}, {Partini}, {Pasanen}, {Pauss},
  {Perez-Torres}, {Persic}, {Peruzzo}, {Pilia}, {Pochon}, {Prada}, {Prada
  Moroni}, {Prandini}, {Puljak}, {Reichardt}, {Reinthal}, {Rhode}, {Rib{\'o}},
  {Rico}, {R{\"u}gamer}, {Saggion}, {Saito}, {Saito}, {Salvati}, {Satalecka},
  {Scalzotto}, {Scapin}, {Schultz}, {Schweizer}, {Shayduk}, {Shore},
  {Sillanp{\"a}{\"a}}, {Sitarek}, {Sobczynska}, {Spanier}, {Spiro}, {Stamerra},
  {Steinke}, {Storz}, {Strah}, {Suri{\'c}}, {Takalo}, {Takami}, {Tavecchio},
  {Temnikov}, {Terzi{\'c}}, {Tescaro}, {Teshima}, {Thom}, {Tibolla}, {Torres},
  {Treves}, {Vankov}, {Vogler}, {Wagner}, {Weitzel}, {Zabalza}, {Zandanel},
  {Zanin}, {Fornasa}, {Essig}, {Sehgal}, \& {Strigari}}]{aleksic11}
{Aleksi{\'c}}, J. {et~al.} 2011, JCAP, 6, 35

\bibitem[{{Aliu} {et~al.}(2012){Aliu}, {Archambault}, {Arlen}, {Aune},
  {Beilicke}, {Benbow}, {Bouvier}, {Bradbury}, {Buckley}, {Bugaev}, {Byrum},
  {Cannon}, {Cesarini}, {Christiansen}, {Ciupik}, {Collins-Hughes}, {Connolly},
  {Cui}, {Decerprit}, {Dickherber}, {Dumm}, {Errando}, {Falcone}, {Feng},
  {Ferrer}, {Finley}, {Finnegan}, {Fortson}, {Furniss}, {Galante}, {Gall},
  {Godambe}, {Griffin}, {Grube}, {Gyuk}, {Hanna}, {Holder}, {Huan}, {Hughes},
  {Humensky}, {Kaaret}, {Karlsson}, {Kertzman}, {Khassen}, {Kieda},
  {Krawczynski}, {Krennrich}, {Lee}, {Madhavan}, {Maier}, {Majumdar},
  {McArthur}, {McCann}, {Moriarty}, {Mukherjee}, {Ong}, {Orr}, {Otte}, {Park},
  {Perkins}, {Pohl}, {Prokoph}, {Quinn}, {Ragan}, {Reyes}, {Reynolds},
  {Roache}, {Rose}, {Ruppel}, {Saxon}, {Schroedter}, {Sembroski}, {{\c
  S}ent{\"u}rk}, {Skole}, {Smith}, {Staszak}, {Telezhinsky}, {Te{\v s}i{\'c}},
  {Theiling}, {Thibadeau}, {Tsurusaki}, {Varlotta}, {Vassiliev}, {Vincent},
  {Vivier}, {Wagner}, {Wakely}, {Ward}, {Weekes}, {Weinstein}, {Weisgarber},
  {Williams}, \& {Zitzer}}]{aliu12}
{Aliu}, E. {et~al.} 2012, \prd, 85, 062001

\bibitem[{{Ando} \& {Nagai}(2012)}]{ando12}
{Ando}, S. \& {Nagai}, D. 2012, JCAP, 7, 17

\bibitem[{{Atwood} {et~al.}(2009){Atwood}, {Abdo}, {Ackermann}, {Althouse},
  {Anderson}, {Axelsson}, {Baldini}, {Ballet}, {Band}, {Barbiellini}, \&
  et~al.}]{atwood09}
{Atwood}, W.~B. {et~al.} 2009, \apj, 697, 1071

\bibitem[{{Baxter} \& {Dodelson}(2011)}]{baxter11}
{Baxter}, E.~J. \& {Dodelson}, S. 2011, \prd, 83, 123516

\bibitem[{{Beck}(2011)}]{beck11}
{Beck}, R. 2011, ArXiv e-prints

\bibitem[{{Beck} \& {Krause}(2005)}]{beck05}
{Beck}, R. \& {Krause}, M. 2005, Astronomische Nachrichten, 326, 414

\bibitem[{{Becker} {et~al.}(1995){Becker}, {White}, \& {Helfand}}]{becker95}
{Becker}, R.~H., {White}, R.~L., \& {Helfand}, D.~J. 1995, \apj, 450, 559

\bibitem[{{Bennett} {et~al.}(2003){Bennett}, {Halpern}, {Hinshaw}, {Jarosik},
  {Kogut}, {Limon}, {Meyer}, {Page}, {Spergel}, {Tucker}, {Wollack}, {Wright},
  {Barnes}, {Greason}, {Hill}, {Komatsu}, {Nolta}, {Odegard}, {Peiris},
  {Verde}, \& {Weiland}}]{bennett03}
{Bennett}, C.~L. {et~al.} 2003, \apjs, 148, 1

\bibitem[{{Bergstr{\"o}m}(2000)}]{bergstrom00}
{Bergstr{\"o}m}, L. 2000, Reports on Progress in Physics, 63, 793

\bibitem[{{Bergstr{\"o}m} {et~al.}(2006){Bergstr{\"o}m}, {Fairbairn}, \&
  {Pieri}}]{bergstrom06}
{Bergstr{\"o}m}, L., {Fairbairn}, M., \& {Pieri}, L. 2006, \prd, 74, 123515

\bibitem[{{Borriello} {et~al.}(2010){Borriello}, {Longo}, {Miele}, {Paolillo},
  {Siffert}, {Tabatabaei}, \& {Beck}}]{borriello10}
{Borriello}, E., {Longo}, G., {Miele}, G., {Paolillo}, M., {Siffert}, B.~B.,
  {Tabatabaei}, F.~S., \& {Beck}, R. 2010, \apjl, 709, L32

\bibitem[{{Calura} {et~al.}(2008){Calura}, {Lanfranchi}, \&
  {Matteucci}}]{calura08}
{Calura}, F., {Lanfranchi}, G.~A., \& {Matteucci}, F. 2008, \aap, 484, 107

\bibitem[{{Charbonnier} {et~al.}(2011){Charbonnier}, {Combet}, {Daniel},
  {Funk}, {Hinton}, {Maurin}, {Power}, {Read}, {Sarkar}, {Walker}, \&
  {Wilkinson}}]{charbonnier11}
{Charbonnier}, A. {et~al.} 2011, ArXiv e-prints

\bibitem[{{Chy{\.z}y} {et~al.}(2000){Chy{\.z}y}, {Beck}, {Kohle}, {Klein}, \&
  {Urbanik}}]{chyzy00}
{Chy{\.z}y}, K.~T., {Beck}, R., {Kohle}, S., {Klein}, U., \& {Urbanik}, M.
  2000, \aap, 355, 128

\bibitem[{{Chy{\.z}y} {et~al.}(2003){Chy{\.z}y}, {Knapik}, {Bomans}, {Klein},
  {Beck}, {Soida}, \& {Urbanik}}]{chyzy03}
{Chy{\.z}y}, K.~T., {Knapik}, J., {Bomans}, D.~J., {Klein}, U., {Beck}, R.,
  {Soida}, M., \& {Urbanik}, M. 2003, \aap, 405, 513

\bibitem[{{Chy{\.z}y} {et~al.}(2011){Chy{\.z}y}, {We{\.z}gowiec}, {Beck}, \&
  {Bomans}}]{chyzy11}
{Chy{\.z}y}, K.~T., {We{\.z}gowiec}, M., {Beck}, R., \& {Bomans}, D.~J. 2011,
  \aap, 529, A94+

\bibitem[{{Cirelli} {et~al.}(2010){Cirelli}, {Panci}, \& {Serpico}}]{cirelli10}
{Cirelli}, M., {Panci}, P., \& {Serpico}, P.~D. 2010, Nuclear Physics B, 840,
  284

\bibitem[{{Colafrancesco} {et~al.}(2006){Colafrancesco}, {Profumo}, \&
  {Ullio}}]{colafrancesco06}
{Colafrancesco}, S., {Profumo}, S., \& {Ullio}, P. 2006, \aap, 455, 21

\bibitem[{{Colafrancesco} {et~al.}(2007){Colafrancesco}, {Profumo}, \&
  {Ullio}}]{colafrancesco07}
---. 2007, \prd, 75, 023513

\bibitem[{{Condon}(1974)}]{condon74}
{Condon}, J.~J. 1974, \apj, 188, 279

\bibitem[{{Condon} {et~al.}(1998){Condon}, {Cotton}, {Greisen}, {Yin},
  {Perley}, {Taylor}, \& {Broderick}}]{condon98}
{Condon}, J.~J., {Cotton}, W.~D., {Greisen}, E.~W., {Yin}, Q.~F., {Perley},
  R.~A., {Taylor}, G.~B., \& {Broderick}, J.~J. 1998, \aj, 115, 1693

\bibitem[{{Crocker} {et~al.}(2010){Crocker}, {Bell}, {Bal{\'a}zs}, \&
  {Jones}}]{crocker10}
{Crocker}, R.~M., {Bell}, N.~F., {Bal{\'a}zs}, C., \& {Jones}, D.~I. 2010,
  \prd, 81, 063516

\bibitem[{{Croft} {et~al.}(2010){Croft}, {Bower}, {Ackermann}, {Atkinson},
  {Backer}, {Backus}, {Barott}, {Bauermeister}, {Blitz}, {Bock}, {Bradford},
  {Cheng}, {Cork}, {Davis}, {DeBoer}, {Dexter}, {Dreher}, {Engargiola},
  {Fields}, {Fleming}, {Forster}, {Gutierrez-Kraybill}, {Harp}, {Helfer},
  {Hull}, {Jordan}, {Jorgensen}, {Keating}, {Kilsdonk}, {Law}, {van Leeuwen},
  {Lugten}, {MacMahon}, {McMahon}, {Milgrome}, {Pierson}, {Randall}, {Ross},
  {Shostak}, {Siemion}, {Smolek}, {Tarter}, {Thornton}, {Urry}, {Vitouchkine},
  {Wadefalk}, {Welch}, {Werthimer}, {Whysong}, {Williams}, \&
  {Wright}}]{croft10}
{Croft}, S. {et~al.} 2010, \apj, 719, 45

\bibitem[{{de Oliveira-Costa} {et~al.}(2008){de Oliveira-Costa}, {Tegmark},
  {Gaensler}, {Jonas}, {Landecker}, \& {Reich}}]{costa08}
{de Oliveira-Costa}, A., {Tegmark}, M., {Gaensler}, B.~M., {Jonas}, J.,
  {Landecker}, T.~L., \& {Reich}, P. 2008, \mnras, 388, 247

\bibitem[{{de Vries} {et~al.}(2004){de Vries}, {Becker}, {White}, \&
  {Helfand}}]{deVries04}
{de Vries}, W.~H., {Becker}, R.~H., {White}, R.~L., \& {Helfand}, D.~J. 2004,
  \aj, 127, 2565

\bibitem[{{Dicker} {et~al.}(2009){Dicker}, {Mason}, {Korngut}, {Cotton},
  {Compi{\`e}gne}, {Devlin}, {Martin}, {Ade}, {Benford}, {Irwin}, {Maddalena},
  {McMullin}, {Shepherd}, {Sievers}, {Staguhn}, \& {Tucker}}]{dicker09}
{Dicker}, S.~R. {et~al.} 2009, \apj, 705, 226

\bibitem[{{Dobler} {et~al.}(2011){Dobler}, {Cholis}, \& {Weiner}}]{dobler11}
{Dobler}, G., {Cholis}, I., \& {Weiner}, N. 2011, \apj, 741, 25

\bibitem[{{Dobler} {et~al.}(2010){Dobler}, {Finkbeiner}, {Cholis}, {Slatyer},
  \& {Weiner}}]{dobler10}
{Dobler}, G., {Finkbeiner}, D.~P., {Cholis}, I., {Slatyer}, T., \& {Weiner}, N.
  2010, \apj, 717, 825

\bibitem[{{Donato} {et~al.}(2004){Donato}, {Fornengo}, {Maurin}, {Salati}, \&
  {Taillet}}]{donato04}
{Donato}, F., {Fornengo}, N., {Maurin}, D., {Salati}, P., \& {Taillet}, R.
  2004, \prd, 69, 063501

\bibitem[{{Feng}(2010)}]{feng10}
{Feng}, J.~L. 2010, \araa, 48, 495

\bibitem[{{Fomalont} \& {Geldzahler}(1979)}]{fomalont79}
{Fomalont}, E.~B. \& {Geldzahler}, B.~J. 1979, \aj, 84, 12

\bibitem[{{Fornengo} {et~al.}(2012){Fornengo}, {Lineros}, {Regis}, \&
  {Taoso}}]{fornengo12}
{Fornengo}, N., {Lineros}, R.~A., {Regis}, M., \& {Taoso}, M. 2012, JCAP, 1, 5

\bibitem[{{Gaensler} {et~al.}(2005){Gaensler}, {Haverkorn}, {Staveley-Smith},
  {Dickey}, {McClure-Griffiths}, {Dickel}, \& {Wolleben}}]{gaensler05}
{Gaensler}, B.~M., {Haverkorn}, M., {Staveley-Smith}, L., {Dickey}, J.~M.,
  {McClure-Griffiths}, N.~M., {Dickel}, J.~R., \& {Wolleben}, M. 2005, Science,
  307, 1610

\bibitem[{{Gallagher} {et~al.}(2003){Gallagher}, {Madsen}, {Reynolds},
  {Grebel}, \& {Smecker-Hane}}]{gallagher03}
{Gallagher}, J.~S., {Madsen}, G.~J., {Reynolds}, R.~J., {Grebel}, E.~K., \&
  {Smecker-Hane}, T.~A. 2003, \apj, 588, 326

\bibitem[{{Geringer-Sameth} \& {Koushiappas}(2011)}]{geringer-sameth11}
{Geringer-Sameth}, A. \& {Koushiappas}, S.~M. 2011, Physical Review Letters,
  107, 241303

\bibitem[{{Gregorini} {et~al.}(1986){Gregorini}, {Ficarra}, \&
  {Padrielli}}]{gregorini86}
{Gregorini}, L., {Ficarra}, A., \& {Padrielli}, L. 1986, \aap, 168, 25

\bibitem[{{Greisen}(2003)}]{greisen03}
{Greisen}, E.~W. 2003, Information Handling in Astronomy - Historical Vistas,
  285, 109

\bibitem[{{Han} {et~al.}(2012){Han}, {Frenk}, {Eke}, {Gao}, {White},
  {Boyarsky}, {Malyshev}, \& {Ruchayskiy}}]{han12}
{Han}, J., {Frenk}, C.~S., {Eke}, V.~R., {Gao}, L., {White}, S.~D.~M.,
  {Boyarsky}, A., {Malyshev}, D., \& {Ruchayskiy}, O. 2012, \mnras, 427, 1651

\bibitem[{{Haslam} {et~al.}(1982){Haslam}, {Salter}, {Stoffel}, \&
  {Wilson}}]{haslam82}
{Haslam}, C.~G.~T., {Salter}, C.~J., {Stoffel}, H., \& {Wilson}, W.~E. 1982,
  \aaps, 47, 1

\bibitem[{{Hooper}(2008)}]{hooper08}
{Hooper}, D. 2008, \prd, 77, 123523

\bibitem[{{Hooper} {et~al.}(2012){Hooper}, {Belikov}, {Jeltema}, {Linden},
  {Profumo}, \& {Slatyer}}]{hooper12}
{Hooper}, D., {Belikov}, A.~V., {Jeltema}, T.~E., {Linden}, T., {Profumo}, S.,
  \& {Slatyer}, T.~R. 2012, \prd, 86, 103003

\bibitem[{{Hooper} {et~al.}(2007){Hooper}, {Finkbeiner}, \&
  {Dobler}}]{hooper07}
{Hooper}, D., {Finkbeiner}, D.~P., \& {Dobler}, G. 2007, \prd, 76, 083012

\bibitem[{{Inoue}(2011)}]{inoue11}
{Inoue}, Y. 2011, \apj, 733, 66

\bibitem[{{Jeltema} \& {Profumo}(2008)}]{jeltema08}
{Jeltema}, T.~E. \& {Profumo}, S. 2008, \apj, 686, 1045

\bibitem[{{Jungman} {et~al.}(1996){Jungman}, {Kamionkowski}, \&
  {Griest}}]{jungman96}
{Jungman}, G., {Kamionkowski}, M., \& {Griest}, K. 1996, \physrep, 267, 195

\bibitem[{{Kepley} {et~al.}(2011){Kepley}, {Zweibel}, {Wilcots}, {Johnson}, \&
  {Robishaw}}]{kepley11}
{Kepley}, A., {Zweibel}, E., {Wilcots}, E., {Johnson}, K., \& {Robishaw}, T.
  2011, ArXiv e-prints

\bibitem[{{Kepley} {et~al.}(2010){Kepley}, {M{\"u}hle}, {Everett}, {Zweibel},
  {Wilcots}, \& {Klein}}]{kepley10}
{Kepley}, A.~A., {M{\"u}hle}, S., {Everett}, J., {Zweibel}, E.~G., {Wilcots},
  E.~M., \& {Klein}, U. 2010, \apj, 712, 536

\bibitem[{{Klein} {et~al.}(1989){Klein}, {Wielebinski}, {Haynes}, \&
  {Malin}}]{klein89}
{Klein}, U., {Wielebinski}, R., {Haynes}, R.~F., \& {Malin}, D.~F. 1989, \aap,
  211, 280

\bibitem[{{Komatsu} {et~al.}(2011){Komatsu}, {Smith}, {Dunkley}, {Bennett},
  {Gold}, {Hinshaw}, {Jarosik}, {Larson}, {Nolta}, {Page}, {Spergel},
  {Halpern}, {Hill}, {Kogut}, {Limon}, {Meyer}, {Odegard}, {Tucker}, {Weiland},
  {Wollack}, \& {Wright}}]{komatsu10}
{Komatsu}, E. {et~al.} 2011, \apjs, 192, 18

\bibitem[{{Laha} {et~al.}(2012){Laha}, {Ng}, {Dasgupta}, \&
  {Horiuchi}}]{laha12}
{Laha}, R., {Ng}, K.~C.~Y., {Dasgupta}, B., \& {Horiuchi}, S. 2012, ArXiv
  e-prints

\bibitem[{{Linden} {et~al.}(2010){Linden}, {Profumo}, \& {Anderson}}]{linden10}
{Linden}, T., {Profumo}, S., \& {Anderson}, B. 2010, \prd, 82, 063529

\bibitem[{{Martin} {et~al.}(2008){Martin}, {de Jong}, \& {Rix}}]{martin08}
{Martin}, N.~F., {de Jong}, J.~T.~A., \& {Rix}, H.-W. 2008, \apj, 684, 1075

\bibitem[{{Mateo}(1998)}]{mateo98}
{Mateo}, M.~L. 1998, \araa, 36, 435

\bibitem[{{Matsumura} {et~al.}(2009){Matsumura}, {Niinuma}, {Kuniyoshi},
  {Takefuji}, {Asuma}, {Daishido}, {Kida}, {Tanaka}, {Aoki}, {Ishikawa},
  {Hirano}, \& {Nakagawa}}]{matsumura09}
{Matsumura}, N. {et~al.} 2009, \aj, 138, 787

\bibitem[{{Maurin} {et~al.}(2001){Maurin}, {Donato}, {Taillet}, \&
  {Salati}}]{maurin01}
{Maurin}, D., {Donato}, F., {Taillet}, R., \& {Salati}, P. 2001, \apj, 555, 585

\bibitem[{{Mayer}(2010)}]{mayer10}
{Mayer}, L. 2010, Advances in Astronomy, 2010

\bibitem[{{McConnachie}(2012)}]{mcconnachie12}
{McConnachie}, A.~W. 2012, \aj, 144, 4

\bibitem[{{Mu{\~n}oz} {et~al.}(2010){Mu{\~n}oz}, {Geha}, \&
  {Willman}}]{munoz10}
{Mu{\~n}oz}, R.~R., {Geha}, M., \& {Willman}, B. 2010, \aj, 140, 138

\bibitem[{{Oppermann} {et~al.}(2012){Oppermann}, {Junklewitz}, {Robbers},
  {Bell}, {En{\ss}lin}, {Bonafede}, {Braun}, {Brown}, {Clarke}, {Feain},
  {Gaensler}, {Hammond}, {Harvey-Smith}, {Heald}, {Johnston-Hollitt}, {Klein},
  {Kronberg}, {Mao}, {McClure-Griffiths}, {O'Sullivan}, {Pratley}, {Robishaw},
  {Roy}, {Schnitzeler}, {Sotomayor-Beltran}, {Stevens}, {Stil}, {Sunstrum},
  {Tanna}, {Taylor}, \& {Van Eck}}]{oppermann12}
{Oppermann}, N. {et~al.} 2012, \aap, 542, A93

\bibitem[{{Papucci} \& {Strumia}(2010)}]{papucci10}
{Papucci}, M. \& {Strumia}, A. 2010, JCAP, 3, 14

\bibitem[{{P{\'e}rez-Torres} {et~al.}(2009){P{\'e}rez-Torres}, {Zandanel},
  {Guerrero}, {Pal}, {Profumo}, {Prada}, \& {Panessa}}]{perez09}
{P{\'e}rez-Torres}, M.~A., {Zandanel}, F., {Guerrero}, M.~A., {Pal}, S.,
  {Profumo}, S., {Prada}, F., \& {Panessa}, F. 2009, \mnras, 396, 2237

\bibitem[{{Planck Collaboration} {et~al.}(2011){Planck Collaboration},
  {Abergel}, {Ade}, {Aghanim}, {Arnaud}, {Ashdown}, {Aumont}, {Baccigalupi},
  {Balbi}, {Banday}, {Barreiro}, {Bartlett}, {Battaner}, {Benabed},
  {Beno{\^i}t}, {Bernard}, {Bersanelli}, {Bhatia}, {Bock}, {Bonaldi}, {Bond},
  {Borrill}, {Bouchet}, {Boulanger}, {Bucher}, {Burigana}, {Cabella},
  {Cardoso}, {Catalano}, {Cay{\'o}n}, {Challinor}, {Chamballu}, {Chiang},
  {Chiang}, {Christensen}, {Colombi}, {Couchot}, {Coulais}, {Crill}, {Cuttaia},
  {Dame}, {Danese}, {Davies}, {Davis}, {de Bernardis}, {de Gasperis}, {de
  Rosa}, {de Zotti}, {Delabrouille}, {Delouis}, {D{\'e}sert}, {Dickinson},
  {Donzelli}, {Dor{\'e}}, {D{\"o}rl}, {Douspis}, {Dupac}, {Efstathiou},
  {En{\ss}lin}, {Finelli}, {Forni}, {Frailis}, {Franceschi}, {Galeotta},
  {Ganga}, {Giard}, {Giardino}, {Giraud-H{\'e}raud}, {Gonz{\'a}lez-Nuevo},
  {G{\'o}rski}, {Gratton}, {Gregorio}, {Grenier}, {Gruppuso}, {Hansen},
  {Harrison}, {Henrot-Versill{\'e}}, {Herranz}, {Hildebrandt}, {Hivon},
  {Hobson}, {Holmes}, {Hovest}, {Hoyland}, {Huffenberger}, {Jaffe}, {Jaffe},
  {Jones}, {Juvela}, {Keih{\"a}nen}, {Keskitalo}, {Kisner}, {Kneissl}, {Knox},
  {Kurki-Suonio}, {Lagache}, {L{\"a}hteenm{\"a}ki}, {Lamarre}, {Lasenby},
  {Laureijs}, {Lawrence}, {Leach}, {Leonardi}, {Leroy}, {Lilje},
  {Linden-V{\o}rnle}, {L{\'o}pez-Caniego}, {Lubin}, {Mac{\'{\i}}as-P{\'e}rez},
  {MacTavish}, {Maffei}, {Mandolesi}, {Mann}, {Maris}, {Marshall},
  {Mart{\'{\i}}nez-Gonz{\'a}lez}, {Masi}, {Matarrese}, {Matthai}, {Mazzotta},
  {McGehee}, {Meinhold}, {Melchiorri}, {Mendes}, {Mennella},
  {Miville-Desch{\^e}nes}, {Moneti}, {Montier}, {Morgante}, {Mortlock},
  {Munshi}, {Murphy}, {Naselsky}, {Natoli}, {Netterfield},
  {N{\o}rgaard-Nielsen}, {Noviello}, {Novikov}, {Novikov}, {Osborne}, {Pajot},
  {Paladini}, {Pasian}, {Patanchon}, {Perdereau}, {Perotto}, {Perrotta},
  {Piacentini}, {Piat}, {Plaszczynski}, {Pointecouteau}, {Polenta}, {Ponthieu},
  {Poutanen}, {Pr{\'e}zeau}, {Prunet}, {Puget}, {Rachen}, {Reach}, {Rebolo},
  {Reich}, {Renault}, {Ricciardi}, {Riller}, {Ristorcelli}, {Rocha}, {Rosset},
  {Rubi{\~n}o-Mart{\'{\i}}n}, {Rusholme}, {Sandri}, {Santos}, {Savini},
  {Scott}, {Seiffert}, {Shellard}, {Smoot}, {Starck}, {Stivoli}, {Stolyarov},
  {Stompor}, {Sudiwala}, {Sygnet}, {Tauber}, {Terenzi}, {Toffolatti}, {Tomasi},
  {Torre}, {Tristram}, {Tuovinen}, {Umana}, {Valenziano}, {Varis}, {Vielva},
  {Villa}, {Vittorio}, {Wade}, {Wandelt}, {Wilkinson}, {Ysard}, {Yvon},
  {Zacchei}, \& {Zonca}}]{planckCIB}
{Planck Collaboration} {et~al.} 2011, \aap, 536, A21

\bibitem[{{Porter} {et~al.}(2011){Porter}, {Johnson}, \& {Graham}}]{porter11}
{Porter}, T.~A., {Johnson}, R.~P., \& {Graham}, P.~W. 2011, ArXiv e-prints

\bibitem[{{Profumo} \& {Ullio}(2010)}]{profumo10}
{Profumo}, S. \& {Ullio}, P. 2010, ArXiv e-prints

\bibitem[{{Siffert} {et~al.}(2011){Siffert}, {Limone}, {Borriello}, {Longo}, \&
  {Miele}}]{siffert11}
{Siffert}, B.~B., {Limone}, A., {Borriello}, E., {Longo}, G., \& {Miele}, G.
  2011, \mnras, 410, 2463

\bibitem[{{Storm} {et~al.}(2012){Storm}, {Jeltema}, {Profumo}, \&
  {Rudnick}}]{storm12}
{Storm}, E., {Jeltema}, T.~E., {Profumo}, S., \& {Rudnick}, L. 2012, ArXiv
  e-prints

\bibitem[{{Strigari}(2012)}]{strigari12}
{Strigari}, L.~E. 2012, ArXiv e-prints

\bibitem[{{Strigari} {et~al.}(2007){Strigari}, {Koushiappas}, {Bullock}, \&
  {Kaplinghat}}]{strigari07}
{Strigari}, L.~E., {Koushiappas}, S.~M., {Bullock}, J.~S., \& {Kaplinghat}, M.
  2007, \prd, 75, 083526

\bibitem[{{Strigari} {et~al.}(2008){Strigari}, {Koushiappas}, {Bullock},
  {Kaplinghat}, {Simon}, {Geha}, \& {Willman}}]{strigari08}
{Strigari}, L.~E., {Koushiappas}, S.~M., {Bullock}, J.~S., {Kaplinghat}, M.,
  {Simon}, J.~D., {Geha}, M., \& {Willman}, B. 2008, \apj, 678, 614

\bibitem[{{Su} {et~al.}(2010){Su}, {Slatyer}, \& {Finkbeiner}}]{su10}
{Su}, M., {Slatyer}, T.~R., \& {Finkbeiner}, D.~P. 2010, \apj, 724, 1044

\bibitem[{{Tasitsiomi} {et~al.}(2004){Tasitsiomi}, {Gaskins}, \&
  {Olinto}}]{tasitsiomi04}
{Tasitsiomi}, A., {Gaskins}, J., \& {Olinto}, A.~V. 2004, Astroparticle
  Physics, 21, 637

\bibitem[{{Walker} {et~al.}(2011){Walker}, {Combet}, {Hinton}, {Maurin}, \&
  {Wilkinson}}]{walker11}
{Walker}, M.~G., {Combet}, C., {Hinton}, J.~A., {Maurin}, D., \& {Wilkinson},
  M.~I. 2011, \apjl, 733, L46

\bibitem[{{Weisz} {et~al.}(2011){Weisz}, {Dalcanton}, {Williams}, {Gilbert},
  {Skillman}, {Seth}, {Dolphin}, {McQuinn}, {Gogarten}, {Holtzman}, {Rosema},
  {Cole}, {Karachentsev}, \& {Zaritsky}}]{weisz11}
{Weisz}, D.~R. {et~al.} 2011, ArXiv e-prints

\bibitem[{{Weniger}(2012)}]{weniger12}
{Weniger}, C. 2012, JCAP, 8, 7

\bibitem[{{White} {et~al.}(1997){White}, {Becker}, {Helfand}, \&
  {Gregg}}]{white97}
{White}, R.~L., {Becker}, R.~H., {Helfand}, D.~J., \& {Gregg}, M.~D. 1997,
  \apj, 475, 479

\end{thebibliography}

\end{document}